\documentclass[12pt,preprint]{aastex}

\newcommand{\kms}{km\,s$^{-1}$\,\,}
\newcommand{\hicat}{{\sc Hicat}\,\,}
\newcommand{\hipass}{{\sc Hipass}\,\,}
\newcommand{\hopcat}{{\sc Hopcat}\,\,}
\newcommand{\hi}{{\sc H\,i}\,\,}
\newcommand{\hii}{{\sc H\,ii}\,\,}
\newcommand{\ho}{$H_0$\,\,}
\newcommand{\hubble}{km s$^{-1}$ Mpc$^{-1}$\,\,}
\newcommand{\mhi}{\mbox{$M_{\rm HI}$}\,\,}
\newcommand{\kaiser}{two-dimensional redshift-space correlation function }

\newcommand{\kmsNS}{km\,s$^{-1}$}
\newcommand{\hicatNS}{{\sc Hicat}}
\newcommand{\hipassNS}{{\sc Hipass}}
\newcommand{\hopcatNS}{{\sc Hopcat}}
\newcommand{\hiNS}{{\sc H\,i}}

\newcommand{\hoNS}{$H_0$}

\newcommand{\pep}{\vspace{3mm}\noindent} 
\newcommand{\minipagesize}{17cm} 

\shorttitle{The Weak Clustering of Gas-Rich Galaxies}
\shortauthors{Meyer et al.}
\begin{document}

\title{The Weak Clustering of Gas-Rich Galaxies}

\author{Martin J. Meyer\altaffilmark{1}, Martin A. Zwaan\altaffilmark{2}, Rachel L. Webster\altaffilmark{3}, \\Michael J.I. Brown\altaffilmark{4}, Lister Staveley-Smith\altaffilmark{5}}
\altaffiltext{1}{Space Telescope Science Institute, 3700 San Martin Drive, Baltimore, MD 21218, U.S.A; martinm@stsci.edu}
\altaffiltext{2}{European Southern Observatory, Karl-Schwarzschild-Str. 2, 85748 Garching b. M{\"u}nchen, Germany; mzwaan@eso.org}
\altaffiltext{3}{School of Physics, The University of Melbourne, VIC 3010, Australia}
\altaffiltext{4}{Princeton University Observatory, Peyton Hall, Princeton, NJ 08544-1001, U.S.A.}
\altaffiltext{5}{Australia Telescope National Facility, CSIRO, P.O. Box 76, Epping, NSW 1710, Australia}

\begin{abstract}
We examine the clustering properties of \hiNS-selected galaxies through an analysis of the \hi Parkes All-Sky Survey Catalogue (\hicatNS) two-point correlation function.   Various sub-samples are extracted from this catalogue to study the overall clustering of \hiNS-rich galaxies and its dependence on luminosity, \hi gas mass and rotational velocity.  These samples cover the entire southern sky $\delta < 0^\circ$, containing up to 4,174 galaxies over the radial velocity range $300-12,700$ \kmsNS.  A scale length of $r_0 = 3.45 \pm 0.25\, h^{-1}$Mpc and slope of $\gamma=1.47\pm0.08$ is obtained for the \hiNS-rich galaxy real-space correlation function, making gas-rich galaxies among the most weakly clustered objects known.  \hiNS-selected galaxies also exhibit weaker clustering than optically selected galaxies of comparable luminosities.  Good agreement is found between our results and those of synthetic \hiNS-rich galaxy catalogues generated from the Millennium Run CDM simulation. Bisecting \hicat using different parameter cuts, clustering is found to depend most strongly on rotational velocity and luminosity, while the dependency on \hi mass is marginal.  Splitting the sample around $v_{\rm rot} = 108$ \kmsNS, a scale length of $r_0 =  2.86\pm 0.46 \,h^{-1}$ Mpc is found for galaxies with low rotational velocities compared to $r_0 =  3.96\pm0.33\,h^{-1}$ Mpc for the high rotational velocity sample.
\end{abstract}
 

\section{Introduction}
\label{sec:intro}

The statistical analysis of galaxy clustering provides key information on the cosmological parameters of the universe and the formation and evolution of galaxies. A simple way of parametrizing galaxy clustering is though the two-point correlation function in its various redshift-space, projected and real-space forms \citep[][]{groth1977,davis1982}.  With the advent of large-scale optical spectroscopic surveys such as the 2dF Galaxy Redshift Survey \citep[2dFGRS, ][]{colless2001} and the Sloan Digital Sky Survey \citep[SDSS, ][]{york2000}, the properties of the galaxy distribution are now able to be studied on cosmologically representative scales.  The large sample sizes have also enabled clustering properties to be examined as a detailed function of parameters such as optical luminosity, morphology, star formation activity and color. From these studies it has been found that the strongest clustering is exhibited by galaxies with high luminosities \citep{norberg2001,norberg2002}, red spectral energy distributions \citep{norberg2002,zehavi2002,zehavi2005}, and relatively passive star formation \citep{madgwick2003}.

In this work, we focus on galaxies that are selected not on their stellar content, but on the amount of cold gas they contain.  Optical selection concentrates on the current stellar properties of galaxies, whereas \hi selection identifies galaxies on their potential to form stars. These galaxies represent a population of more slowly evolving galaxies, still having a large fuel reservoir available for conversion into stars.   The \hi Parkes All-Sky Survey (\hipassNS) Catalogue \citep[\hicatNS;][]{meyer2004} provides such a sample over the entire southern sky.  As a blind \hi survey, \hipass is also not biased by extinction, providing a unique view of regions such as the Zone of Avoidance that are difficult to observe in the optical.  

This paper provides a detailed analysis of the \hicat two-point correlation function, building on earlier work by \citet{meyer2004b} and \citet{ryan2006} which also examine the clustering properites of \hipass galaxies.  Here, we present a more detailed analysis, with a strong emphasis on the clustering dependencies on various galaxy parameters.

The basic properties of \hicat are described in Section~\ref{sec:cf_data}, with a discussion of the two-point correlation function technique following in Section~\ref{sec:corr_func_desc}.  Sections~\ref{sec:clustering}~and~\ref{sec:dependencies} present the main results of this work, first covering the redshift-space, projected and real-space correlation functions, followed by an examination of the dependency of galaxy clustering on luminosity, \hi mass, and rotational velocity.  These results are discussed in Section~\ref{sec:discussion} with a summary given in Section~\ref{sec:conclusions}.  A Hubble constant of \ho = 100 \hubble is used throughout to compare results with existing published work.

\section{Data}
\label{sec:cf_data}

The galaxy data are taken from \hicatNS, the largest blind catalogue of \hi sources compiled to date.  \hicat accurately determines the position and redshift of the galaxies simultaneously, one of the unique benefits of an \hi survey.   We provide a brief description of this dataset here, referring the reader to the relevant catalogue and data papers for a full discussion \citep{barnes2001,meyer2004,zwaan2004}.

 \hicat contains 4,315 sources over the southern sky $\delta < +2^\circ$ and spanning the velocity range 300 to 12,700 \kmsNS.   The catalogue was compiled using a combination of automatic and manual procedures from \hipass data.  Observations for this survey were carried out from 1997 to 2000 with the Parkes 64 metre radio telescope.  The dataset has a final spatial resolution of 15.5 arcmin and velocity resolution of 18 \kms following smoothing.  Average noise for the data is 13 mJy beam$^{-1}$, with data at low galactic latitudes having slightly elevated noise levels \citep{zwaan2004}.  The completeness and reliability of the sample was measured using a combination of fake sources added to the data and follow-up observations respectively, and is described in detail in \citet{zwaan2004}.  From this, 99 per cent of inserted sources are retrieved for sources with peak flux $>$ 84 mJy or an integrated flux $>$ 9.4 Jy \kmsNS.  Similarly, 99 per cent of catalogue sources are found to be real for peak fluxes $>$ 58 mJy or integrated flux $>$ 8.2 Jy \kmsNS.  Overall reliability for the entire catalogue is found to be 95 per cent.  To give a feeling of survey depth, \hicat has a complete sampling (integrated flux limited) of $L^*$ galaxies to a distance of $\sim 40\, h^{-1}$ Mpc, although such galaxies are present in the catalogue to $\sim 80\, h^{-1}$Mpc.

\section{The Two-Point Correlation Function}
\label{sec:corr_func_desc}

The two-point correlation function, $\xi$, provides a simple measure
of galaxy clustering.  This is computed by comparing the number of
galaxy pairs at different on-sky and radial separations ($\sigma$ and
$\pi$ respectively) between the real data sample and those of a
randomly generated dataset.  The random dataset is constructed to have
the same selection function and boundaries as the real dataset, but
with an unclustered Poissonian distribution of galaxies.  The \hipass selection function is derived using a stepwise maximum likelihood technique \citep{zwaan2005}.  In
particular, $\xi(\sigma,\pi)$ is defined to give the excess
probability of finding a galaxy pair on the given scale $(\sigma,\pi)$
compared to the random dataset.  A value $\xi(\sigma,\pi)=1$ thus
corresponds to the real dataset having twice the probability of
containing a galaxy pair on the specified scale compared to the random
catalogue.  In this work, the relations $\sigma = [(v_i +
v_j)/H_0]\tan(\theta/2)$ and $\pi = |(v_i - v_j)/H_0|$ are used to
calculate the transverse and radial separations respectively
\citep[following][]{davis1983}.  Velocities for the real dataset are in the
heliocentric frame of reference, which provide a first order
compromise between the Local Group and Cosmic Microwave Background
standard of rest frames spanned by the \hicat sample \citep{meyer2006}.

\subsection{Estimators}
\label{sec:estimators}

Galaxy pair number counts of the two samples can be compared using a
variety of techniques, or estimators.  Historically, three main
estimators have been used: \citet{davis1982}, \citet{hamilton1993},
\citet{landay1993}.  In this work, we use the last of these.  The
\citeauthor{landay1993} estimator minimizes the effects of errors in
the measurement of sample mean galaxy density, as well as problems of
using the observed sample itself to measure the mean density.  This
estimator is given by:

\begin{eqnarray}
\nonumber &\xi_{\rm LS}(\sigma,\pi) = \frac{1}{RR(\sigma,\pi)}\left[DD(\sigma,\pi)\left(\frac{n_R}{n_D}\right)^2\right.\\
&\left. - 2DR(\sigma,\pi)\left(\frac{n_R}{n_D}\right) + RR(\sigma,\pi)\right],
\end{eqnarray}

\pep where $DD(\sigma,\pi)$ is the number of pairs at separation $\sigma$
and $\pi$ in the real data sample, and $DR(\sigma,\pi)$ is the number
of pairs when matching the real data sample with the random catalogue,
and $RR(\sigma,\pi)$ is the number of pairs in the random catalogue at
the specified separations.  The values $n_D$ and $n_R$ are the galaxy
number densities in the data and random samples respectively.  These
are needed to normalize pair counts between the two catalogues, as
the random data sample is generated to contain many more points than the
data sample to reduce statistical variation.

For non-volume limited samples, such as the one used here, an
additional problem that arises is the effect of the sample
selection function.  As noted in \citet{ratcliffe1998}, if no pair
weighting scheme is used, pair counts are dominated by galaxies at the
peak of the survey selection function, effectively reducing the survey
volume.  On the other hand, if an inverse selection function weighting
is used, this causes pairs at high velocities to dominate where galaxy
counts are low.  A compromise is one that minimizes the variance in
the estimate of $\xi$, as discussed in \citet{davis1982} and
\citet{hamilton1993}.  In this case, rather than weighting each pair
count equally (DD, DR and RR summed using $w_{ij} = 1$), the weighting
for a given pair $w_{ij}$ with redshift-space separation
$s=\sqrt{\sigma^2+\pi^2}$ can be calculated by multiplying individual
galaxy weights ($w_{ij} = w_iw_j$) which are calculated from
\citep{efstathiou1988,hawkins2003}:

\begin{equation}
w_{i} = w(r_i,s)=\frac{1}{1+4\pi n_D\phi(r_i)J_3(s)},
\end{equation}

\pep where $\phi(r_i)$ is the survey selection function at the
distance $r_i$ (=$v_i$/\hoNS) of the galaxy under consideration.  For
close galaxies, where the selection function is large, this weighting
has the property $w_i \propto 1/\phi(r_i)$ as desired, whereas at
large distances when the selection function is small $w_i \sim 1$.
$J_3(s)$ is defined by

\begin{equation}
J_3(s) = \int^s_0 s'^2 \xi(s')ds'.
\end{equation} 

\pep This requires the redshift-space correlation function $\xi(s)$, which is one of the quantities we are trying to determine.  However, for the calculation of weights it is sufficient to assume a power-law form $\xi(s)=(s/s_0)^{-\gamma}$ for the calculation of weights, where the values $s_0 = 5.0$ and $\gamma = 1.8$ are used.  Furthermore, $\xi(s)$ is set to zero for values $s > 30$ Mpc \citep[see e.g.][]{fisher1994}.  As noted by a number of authors, results are not sensitive to the exact form of $J_3(s)$ \citep{hawkins2003,ratcliffe1998,zehavi2002}.  To avoid excess noise in the measured correlation functions through the over-weighting of a few distant galaxies, it is found to be necessary to limit the \hicat sample to $v < 6000$ \kms for the weighted samples. This reduces the sample size to 3820 in the weighted analysis.   The galaxy number density, $n_D$, is calculated according to \citep{davis1982,willmer1997}:

\begin{equation}
n_D = \frac{\sum w(r_i,s=30 {\rm Mpc})}{\int dV\phi(r)w(r)}
\end{equation}

\pep where the sum is taken over all galaxies $\phi(r_i) > 0.001$ and
the volume integral equivalently (limiting errors caused by sparse
sampling).  This expression for $n_D$ is circular in definition
(through $w$), but $n_D$ converges rapidly if the expressions are
evaluated iteratively.

The normalizing ratio $n_R/n_D$ can be calculated when using this
weighting scheme using \citep{davis1982,fisher1994}:

\begin{equation}
\frac{n_D}{n_R} = \frac{\sum^{i=N_D}_{i=1}w_i(r_i,s=30 {\rm Mpc})}{\sum^{j=N_R}_{j=1}w_j(r_j,s=30 {\rm Mpc})},
\end{equation}

\pep where $N_D$ and $N_R$ are the number of objects in the real and
random samples respectively.  Results from both the weighted and
unweighted schemes are presented in this work.  Only galaxy pairs with
separations $< 50^{\circ}$ are included in the work presented here
\citep{davis1983}.

Like early optical redshift surveys, \hipass spans a relatively small
volume and has significant structure on scales comparable to that of
the survey region.  Particularly notable are two large-scale structure
features at the peak of the \hicat radial velocity distribution.  This raises
the possibility that the specific location of these structure may
influence our clustering results, i.e. the survey region may not be
truly representative of the homogenous universe on larger scales.
However, the use of a weighting scheme as described aims to maximize
the volume contributing to the measured clustering, while still
utilizing the more significant number counts at closer distances.  Our analysis of synthetic catalogues also indicates that this effect should not significantly bias our results (see Section \ref{sec:Millennium}).

A final effect that may influence \hicat results is source confusion due to the relatively large \hipass beam (15.5 arcmin).  At the applied weighted sample distance limit of 60 $h^{-1}$Mpc, the beam size corresponds to $\sim 0.3\, h^{-1}$Mpc (cf. 0.27 $h^{-1}$Mpc for the smallest separation bins examined here), and less than half that for galaxies at the peak of the \hipass redshift distribution.  

\subsection{Random Samples}
\label{sec:random}

\begin{figure*}
\begin{center}
\includegraphics[trim=1cm 1cm 3cm 2cm,width=7.4cm,keepaspectratio=true]{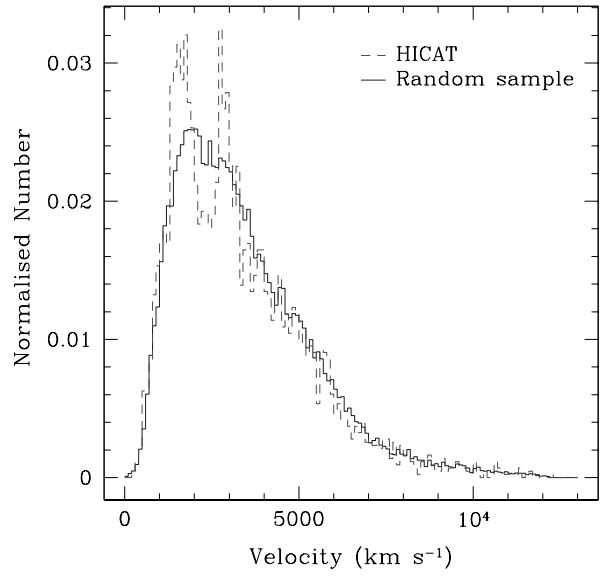}
\end{center}
\caption[Real and random velocity distribution comparisons]{Radial velocity
histogram of the \hicat galaxies (dashed line) compared with the
generated random sample (solid line).}
\label{fig:random}
\end{figure*}

The random samples in this study are generated to match the underlying radial velocity distribution of the real datasets.  This is done using `kernel density estimation,' which finds the optimal Gaussian kernel width that should be used to smooth the data \citep{wand1995}. On-sky positions are random.  The normalised radial velocity distributions of the full catalogue and its random sample are shown in Figure~\ref{fig:random}.  As a check, this method was compared with that of overlaying the \hicat completeness function onto a volume limited random sample of galaxies with \hi masses, peak fluxes and rotational velocities derived from the \hi mass function.   While giving consistent correlation function results with those from the kernel density estimation method, this method was not used due to its model dependence.

\section{HICAT Galaxy Clustering}
\label{sec:clustering}

\subsection{Two-Dimensional Redshift-Space Correlation Function}

\begin{figure*}
\includegraphics[trim=0cm 0cm 2cm 0cm,width=7cm,keepaspectratio=true]{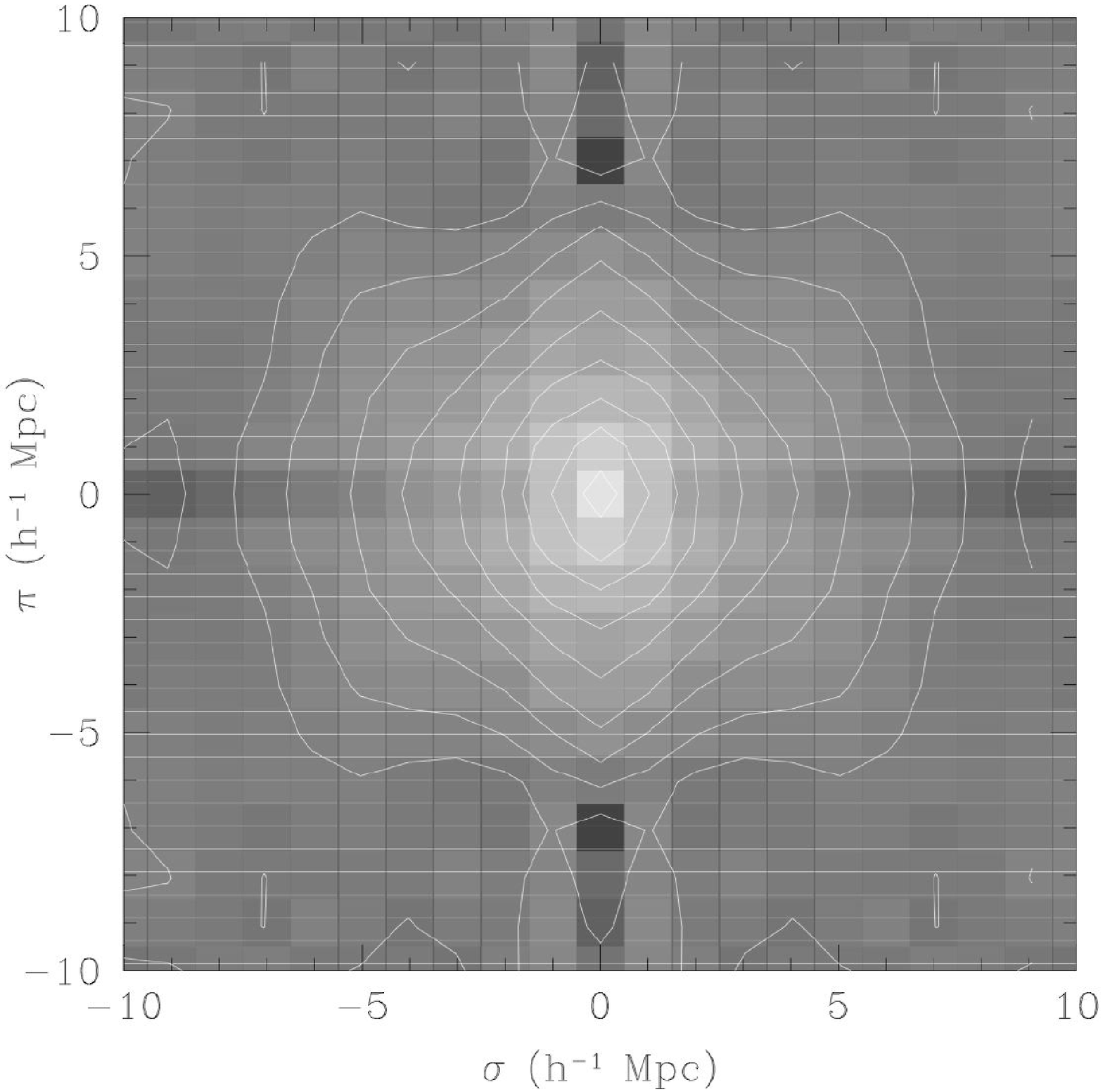}
\includegraphics[trim=-3cm 0cm 5cm 0cm,width=7cm,keepaspectratio=true]{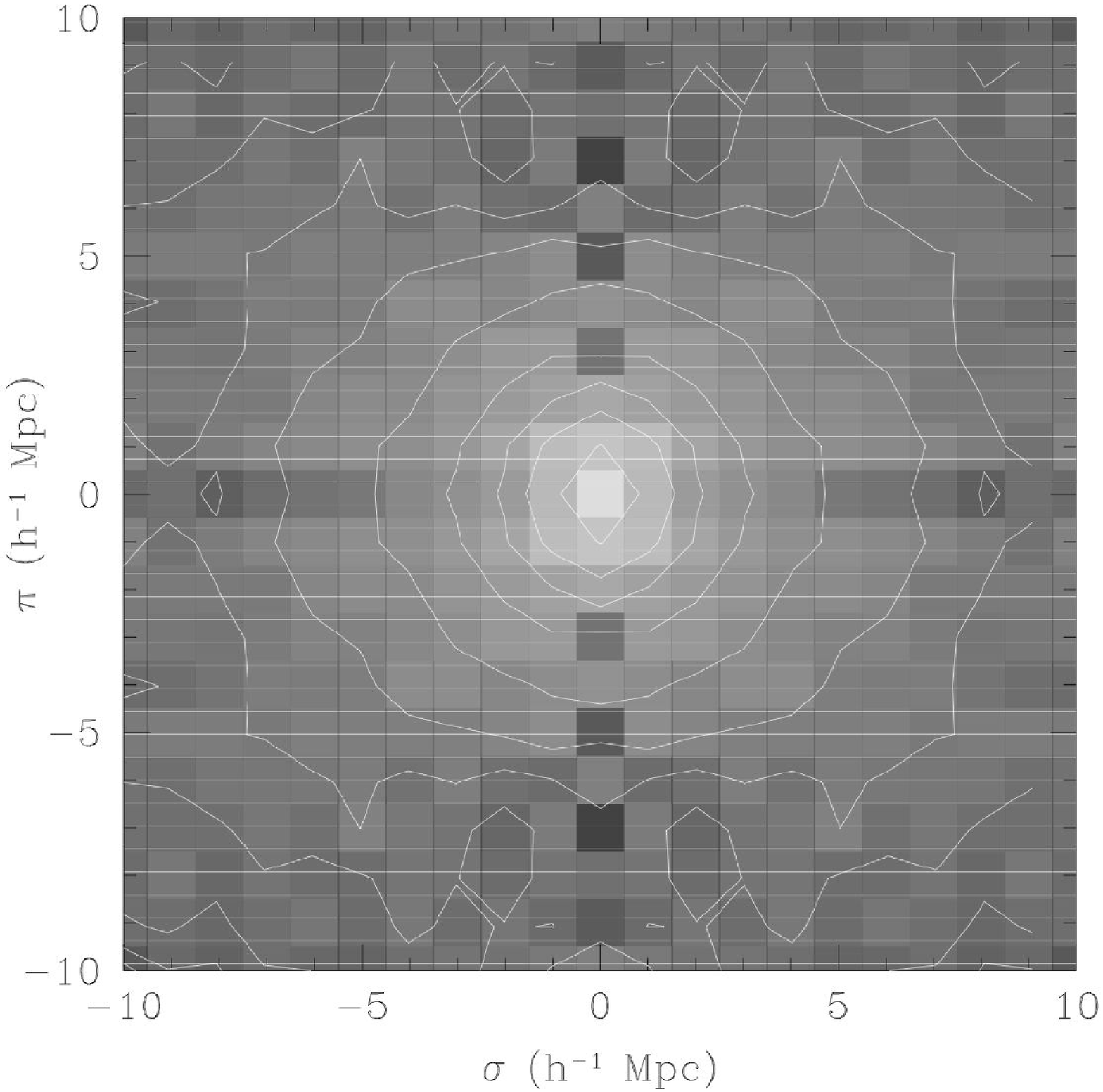}
\caption[\kaiser diagram]{\kaiser diagram for \hicat galaxies:
unweighted (left) and weighted (right).  Lighter shades correspond to
high values of the two-point correlation function.}
\label{fig:kaiser}
\end{figure*}

Figure~\ref{fig:kaiser} plots the \kaiser diagrams for the sample,
showing both the weighted and unweighted versions.  These diagrams
have been created by mirroring the original calculated function into
each of the other three quadrants.  Contours are fitted at logarithmic
intervals to a smoothed version of the correlation function images.

There are two distortions caused by the peculiar velocities of
galaxies that are commonly observed in these diagrams for
optically-selected galaxy samples.  On small angular ($\sigma$)
scales, the correlation function contours are stretched from their
real-space circular shape outward in the radial ($\pi$) direction.
This is the non-linear `Finger-of-God' effect, caused by the
line-of-sight velocity dispersion of galaxies in gravitationally bound
structures such as galaxy groups and clusters.  The
second redshift-space distortion observed is the linear large-scale
flattening of the correlation function contours in the $\pi$
direction, caused by the coherent infall of galaxies into large-scale
over-densities \citep{kaiser1987}.

Both of these effects can be seen in the \hicat sample.  The
Finger-of-God effect is best seen in the unweighted samples, which are
dominated by nearby galaxy pairs, and the large-scale infall is most
apparent in the weighted correlation functions where the catalogue
effective volume is larger.  The line-of-sight velocity dispersion
effects are likely to be due to \hiNS-rich galaxies in less dense
gravitationally bound concentrations such a galaxy groups.  Objects in
larger concentrations, such a galaxy clusters, will contribute less to
the observed dispersion effect than for optically-selected samples
given the relative paucity of \hiNS-rich galaxies in cluster
environments \citep{waugh2002}.  The large-scale coherent infall of
\hiNS-rich galaxies is only sampled on relatively small scales given the
shallow nature of \hicatNS.

\subsection{Redshift-Space Correlation Function}

To compare the correlation results of \hicat to those of optically-selected samples in a single dimension, the first cut on the \kaiser diagram that can be examined is the redshift-space correlation function, $\xi(s)$.  This is constructed by taking the radial average of the two-point correlation function $\xi(\sigma,\pi)$, defining $s=\sqrt{\sigma^2 + \pi^2}$ as before.

Errors are measured using jackknife re-sampling \citep[see][]{lupton1993}, dividing the sample under consideration into 24 RA bins.  The redshift-space correlation function is then re-measured 24 times, each time leaving out one bin of RA.  The error in a given redshift bin $s$ is given by ($N = 24$):

\begin{equation}
\sigma_{\xi(s)}^{2} = \frac{N-1}{N}\sum^{N}_{i=1}(\bar{\xi}(s) - \xi_{i}(s))^2
\end{equation}

\pep Jackknife errors take into account random errors and to some degree those due to cosmic variance, although measurement of these latter errors is limited by the small \hicat effective volume.  We do not determine the full covariance matrix for the binned correlation function data (the bins themselves are not independent) due to the small sample size.  However, Section~\ref{sec:Millennium} provides further analysis of our error estimates through an examination of simulated \hi galaxy catalogues.  Systematic errors are not taken into account.  

\begin{figure*}
\begin{center}
\includegraphics[trim=1cm 0cm 3cm 1cm,width=6.6cm,keepaspectratio=true]{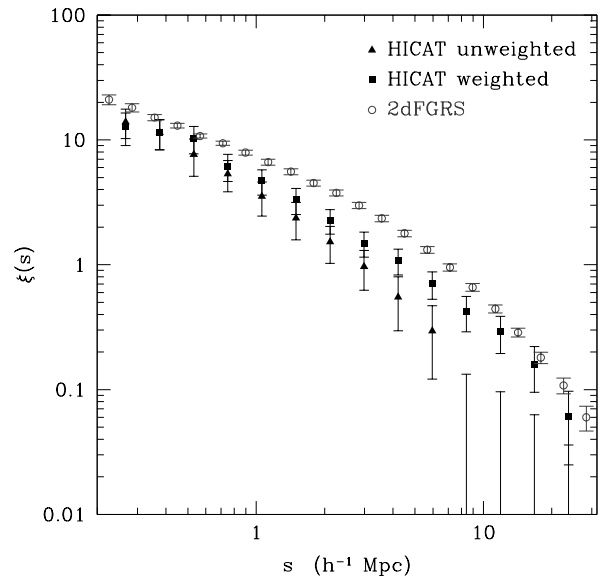}
\end{center}
\caption[Redshift-space correlation functions]{Redshift-space two-point correlation function for \hicat galaxies samples compared with 2dFGRS results. Triangles are the unweighted \hicat results, squares are the weighted \hicat results and circles are those for the
2dFGRS \citep{hawkins2003}. }
\label{fig:cf_projreal}
\end{figure*}

Weighted and unweighted results are given in Figure~\ref{fig:cf_projreal}, with 2dFGRS results also included.  It is clear that the \hicat galaxies are more weakly clustered than the 2dFGRS galaxies on all scales $< 30\, h^{-1}$Mpc.  In the following sections we explore this offset in more detail, using the projected two-point correlation function to obtain the real space correlation function for \hiNS-selected galaxies.

\subsection{Projected Correlation Function}

A difficulty of the redshift-space correlation function is that it is
still affected by the peculiar velocities of galaxies, which may be
different for \hiNS- and optically-selected samples.  To compare the
true spatial clustering of these galaxies, it is necessary to examine
the clustering properties free of redshift-space distortions.  One way
this can be done is through the projected correlation function,
measured by integrating the \kaiser diagram along the $\pi$ axis:

\begin{equation}
\frac{\Xi(\sigma)}{\sigma} = \frac{2}{\sigma}\int^{D_{\rm lim}}_{0}\xi(\sigma,\pi)d\pi.
\label{eqn:xi_proj}
\end{equation}

\begin{figure*}
\begin{minipage}{19cm}
\begin{center}
\includegraphics[trim=10mm 0mm 0mm 10mm,width=6.8cm,keepaspectratio=true]{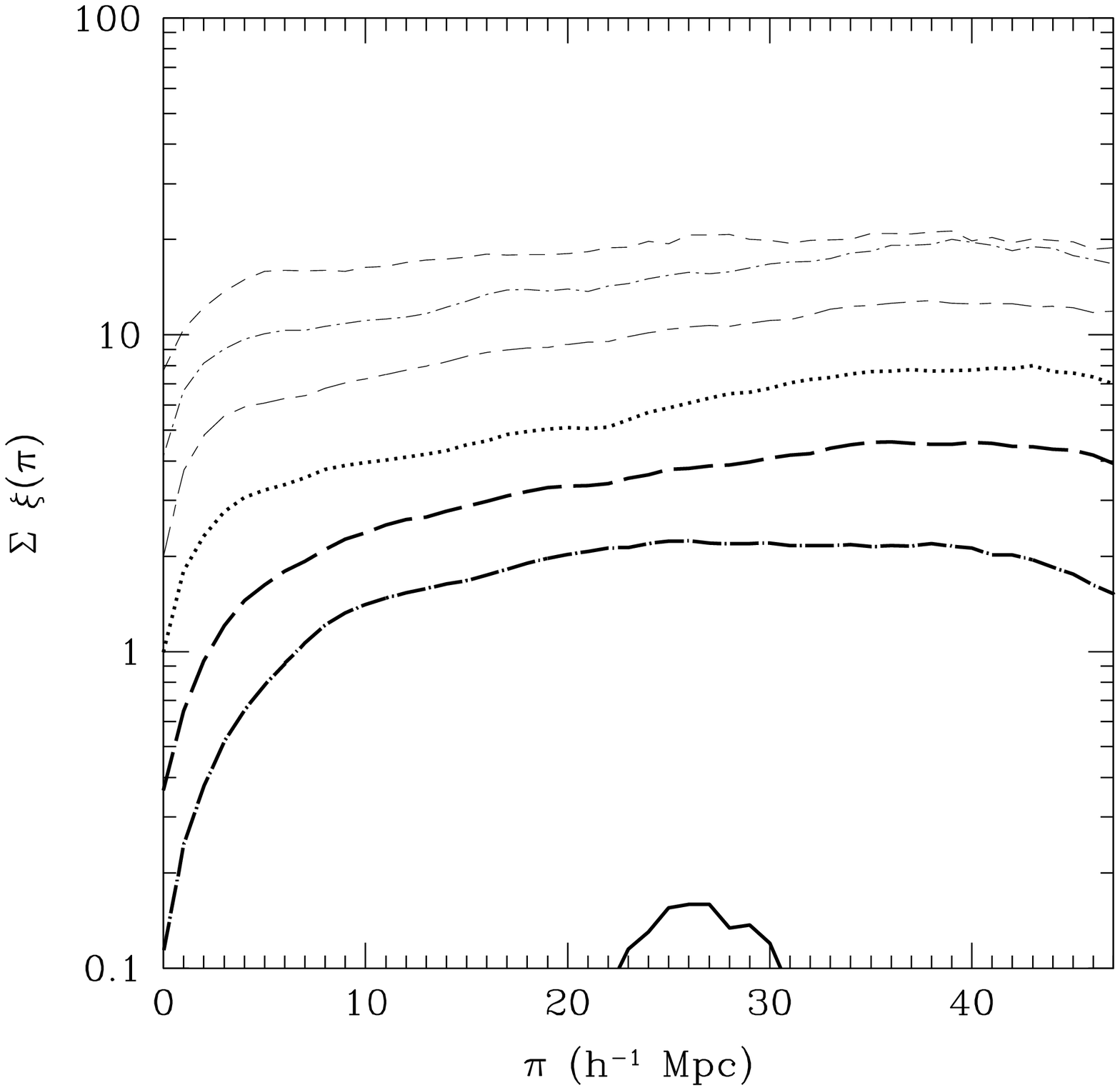}
\includegraphics[trim=10mm 0mm 0mm 10mm,width=6.8cm,keepaspectratio=true]{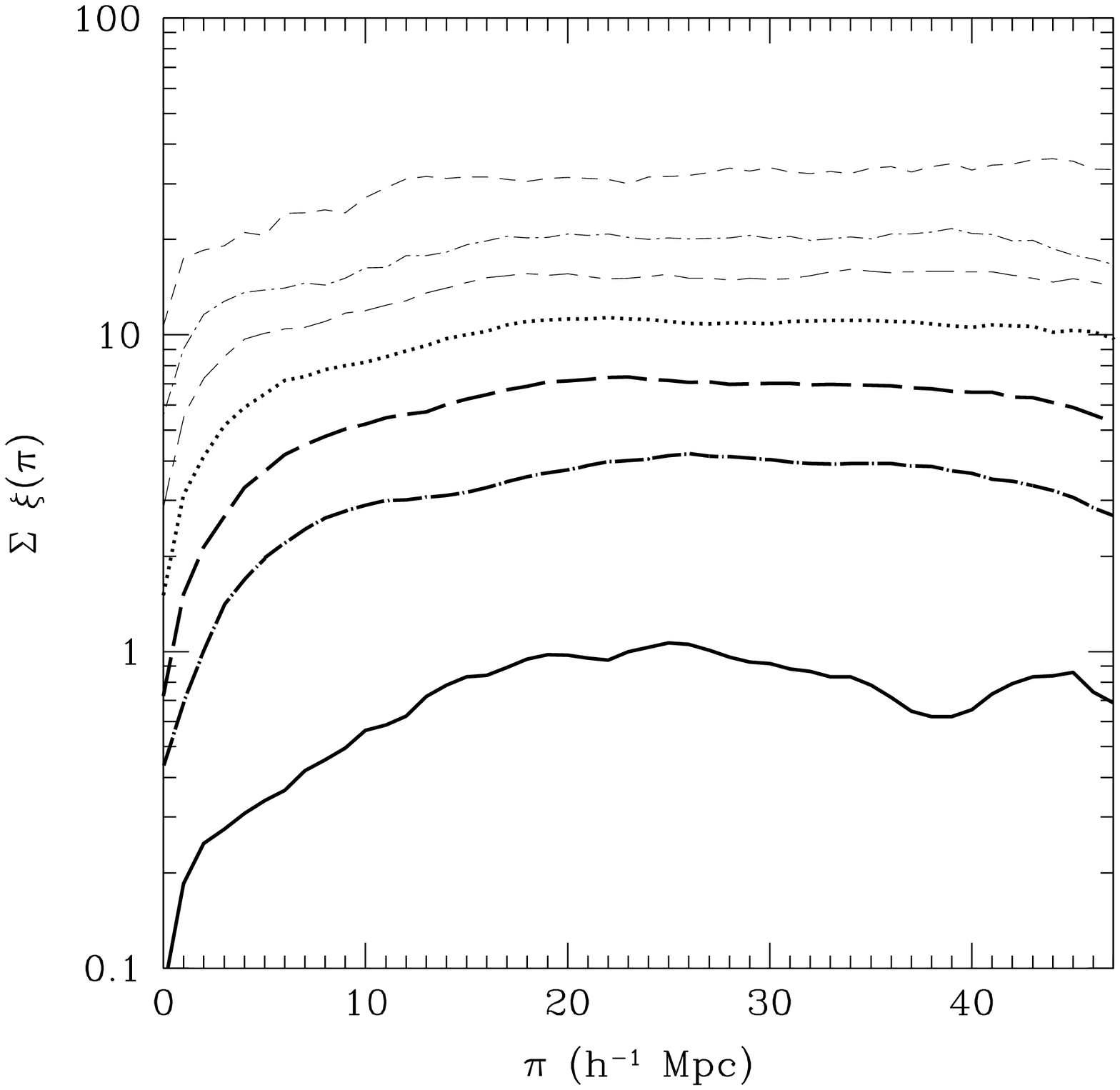}
\includegraphics[trim=4cm 5cm 5cm 0cm,height=6cm,keepaspectratio=true]{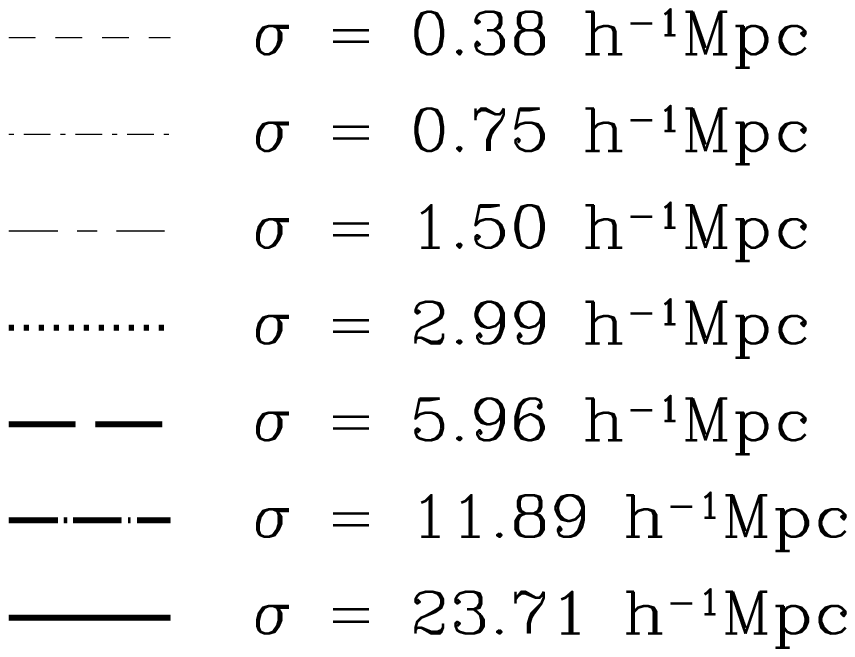}
\end{center}
\end{minipage}
\caption[$\xi_{ls}$ integrals]{Unweighted and weighted $\xi$ integrals
($\int_0^{\pi} \xi(\sigma,\pi')d\pi'$): (left) unweighted, (right)
weighted. For clarity, only every alternate integral is shown.  Each $\sigma$ bin is plotted in a different line style as
specified by the key.  The bin centre values are given in units of $h^{-1}$Mpc.}
\label{fig:XI_sums}
\end{figure*}

\pep The upper limit $D_{\rm lim}$ is chosen here at the point where
the integral converges.  In this work the limit $D_{\rm lim} = 25
h^{-1}$ Mpc is used, with the integrals broadly reaching a plateau at
this point in the weighted samples (see Figure~\ref{fig:XI_sums}). It
should be noted that for the unweighted samples, the integrals do not
completely converge over the range of separations scales probed here.
This point is discussed further in Section~\ref{sec:powerlaw}.
 
\begin{figure*}
\begin{center}
\includegraphics[trim=1cm 0cm 3cm 1cm,width=6.6cm,keepaspectratio=true]{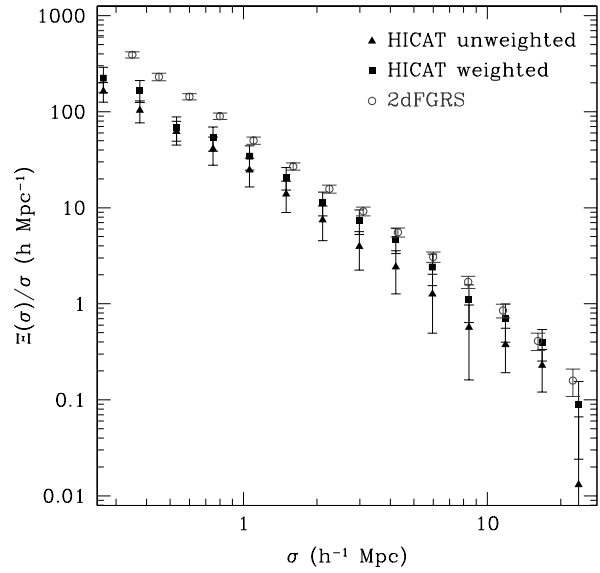}
\end{center}
\caption[Projected real-space correlation functions]{Projected
two-point correlation function for the \hicat galaxy samples compared
with 2dFGRS results.  Triangles are the unweighted \hicat results,
squares are the weighted \hicat results and circles are those for the
2dFGRS \citep{hawkins2003}.}
\label{fig:cf_real}
\end{figure*}

Errors are calculated for the projected correlation functions using
jackknife re-sampling as before.  The resultant correlation functions
are shown in Figure~\ref{fig:cf_real}.  It can be seen that the projected correlation function is much more power-law in shape compared to the redshift-space correlation function.

\subsection{Real-Space Correlation Function}

Two methods are used to obtain the non-projected real-space
correlation function.  First, a power-law form is assumed for the
real-space correlation function, and second the projected
correlation function is inverted to retrieve the real-space
correlation function without this assumption.

\subsubsection{Correlation Function Assuming Power-Law}
\label{sec:powerlaw}

Following \citet{davis1983} and \citet{norberg2001}, if the integral
up to $D_{\rm lim}$ in Equation~\ref{eqn:xi_proj} includes nearly all
correlated pairs, the projected correlation function is related to the
real-space correlation function $\xi(r)$ by:

\begin{equation}
\label{eqn:proj_corr1}
\frac{\Xi(\sigma)}{\sigma} = \frac{2}{\sigma}\int^{\infty}_{\sigma}\xi(r)\frac{r dr}{(r^2 - \sigma^2)^{1/2}}.
\end{equation}

\pep Assuming the real-space correlation function has a power-law form
$\xi(r)=\left(\frac{r}{r_0}\right)^{-\gamma}$, the above integral can
be evaluated in terms of gamma functions giving:

\begin{equation}
\frac{\Xi(\sigma)}{\sigma} =
\left(\frac{r_0}{\sigma}\right)^\gamma\frac{\Gamma(\frac{1}{2})\Gamma(\frac{\gamma-1}{2})}{\Gamma(\frac{\gamma}{2})}
= \left(\frac{r_0}{\sigma}\right)^\gamma A(\gamma).
\end{equation}

\begin{figure*}
\includegraphics[trim=0cm 1cm 2cm 0cm,width=7.2cm,keepaspectratio=true]{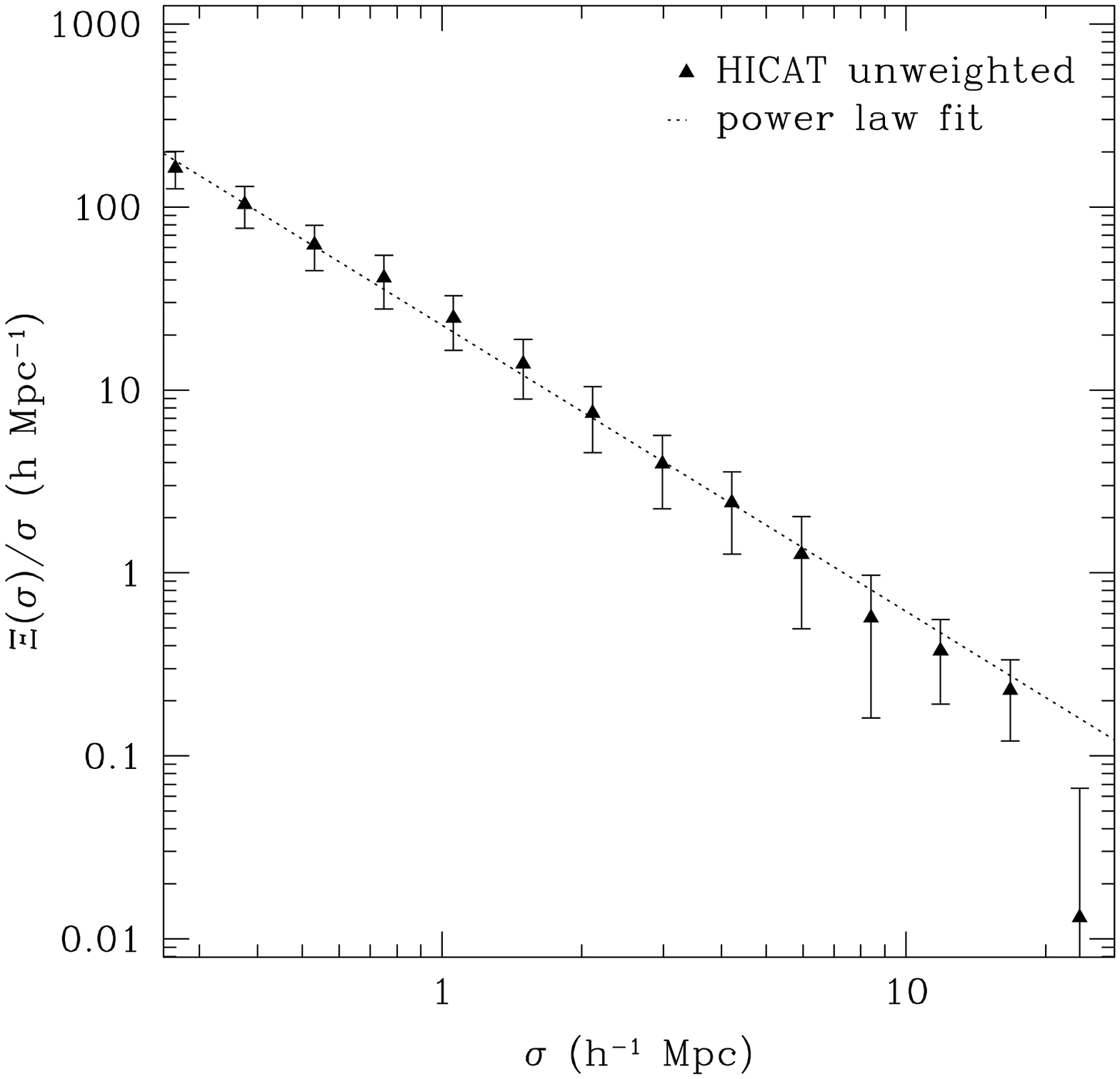}
\includegraphics[trim=-3cm 1cm 5cm 0cm,width=7.2cm,keepaspectratio=true]{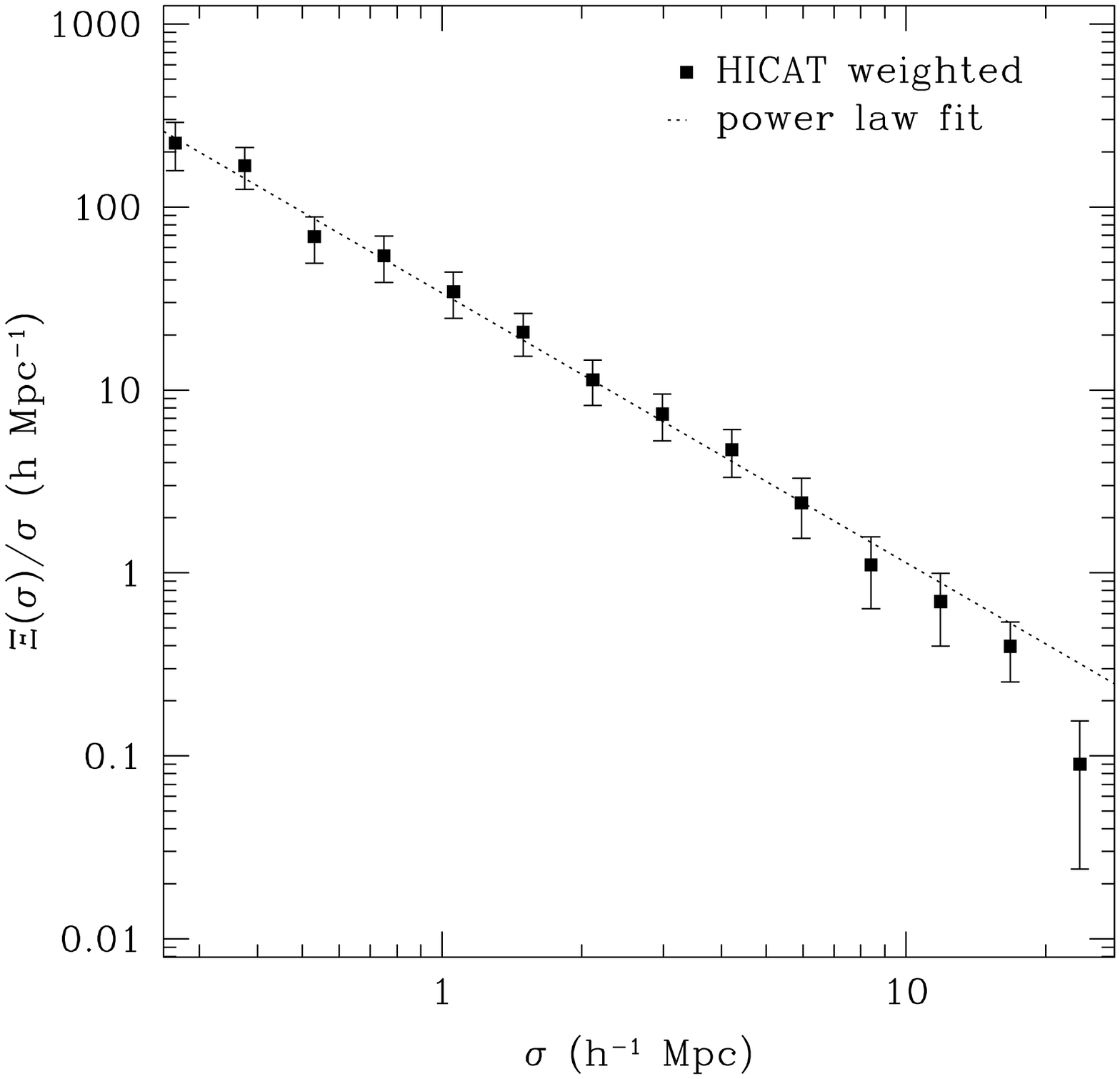}
\caption[Projected correlation function power-law fits]{Projected
real-space correlation functions (points) with power-law fits used to obtain
$r_0$ and $\gamma$: (left) unweighted, (right) weighted. Fitting restricted to points $\sigma<10\, h^{-1}$Mpc. }
\label{fig:cf_real_fit}
\end{figure*}

\pep To calculate $r_0$ and $\gamma$, the general power-law form $\Xi(\sigma) = a_1\sigma^{a_2}$ is fitted to the projected correlation function using the Levenberg-Marquardt nonlinear least-squares method \citep{press1992}.  Resultant fits are shown in Figure~\ref{fig:cf_real_fit}.  Only points $\sigma<10\, h^{-1}$Mpc are used in these fits to restrict the data to the power-law part of the plotted correlation functions. The parameters $r_0$ and $\gamma$ are then given by:

\begin{eqnarray}
r_0 &=& \left(\frac{a_1}{A(1-a_2)}\right)^{\frac{1}{1-a_2}}\label{eqn:r0}\\
\gamma &=& 1-a_2\label{eqn:gamma}
\end{eqnarray}

\pep Errors on $r_0$ and $\gamma$ are calculated taking the square
root of the diagonal elements of the measured covariance matrix as the
errors for $a_1$ and $a_2$ then propagating appropriately.  The
off-diagonal term is also included given $r_0$ and $\gamma$ are not
independent in the fitting process:

\begin{eqnarray}
\nonumber \sigma_{r_0} &=& \left[\left(\frac{\partial r_0}{\partial a_1}\right)^2\!\!\!\sigma_{a_1}^2+\left(\frac{\partial r_0}{\partial a_2}\right)^2\!\!\!\sigma_{a_2}^2 + \right.\\
&&\left.2\left(\frac{\partial r_0}{\partial a_1}\right)\!\!\left(\frac{\partial r_0}{\partial a_2}\right)\!\sigma_{a_1a_2}^2\right]^{1/2}
\end{eqnarray}
\begin{equation}
\sigma_{\gamma} = \left|\frac{\partial \gamma}{\partial a_2}\right|\sigma_{a_2}.
\end{equation}

\pep Final parameter results for the weighted and unweighted samples are given in Table~\ref{tab:r0gamma}.  As discussed earlier, there is still some dependence on the $\pi$ axis integration limit for the projected correlation function in the case of the unweighted sample, and changing this limit from $25\,h^{-1}$Mpc to $35\,h^{-1}$Mpc alters the measured clustering scale length from $2.70\pm0.21\,h^{-1}$Mpc to $3.05\pm0.23\,h^{-1}$Mpc (cf. $3.45\pm0.25\,h^{-1}$Mpc to $3.56\pm0.23\,h^{-1}$Mpc for the weighted sample). Also included for comparison in Table~\ref{tab:r0gamma} are the 2dFGRS results 
from \citet{norberg2002} examining clustering as a function of luminosity and spectral type, and the recent SDSS results of \citet{zehavi2005} also investigating clustering as a function of luminosity. The quoted \citet{norberg2002} results correspond to the strongest and weakest clustered magnitude ranges for both early and late-type galaxies. The SDSS faint and bright results are those at the extreme ends of the measured luminosity distribution. 

From the correlation function parameter values, it can be seen that the \hicat scale lengths are all weaker than the 2dFGRS results, although within errors of faint late-type galaxies. \hicat galaxies have similar $r_0$ values to those of SDSS galaxies with optical luminosities $-19<M_R<-18$ \citep[$r_0 = 3.51\pm 0.32$;][]{zehavi2005}, though less luminous SDSS galaxies have even lower values of $r_0$ (see Table~\ref{tab:r0gamma}). In all cases, the \hicat correlation function exhibits a flatter slope than optically selected samples, reducing the comparative clustering strength of \hicat galaxies on small scales.

\begin{figure*}
\includegraphics[trim=0.5cm 1cm 5cm 4cm,width=7cm,keepaspectratio=true.clip]{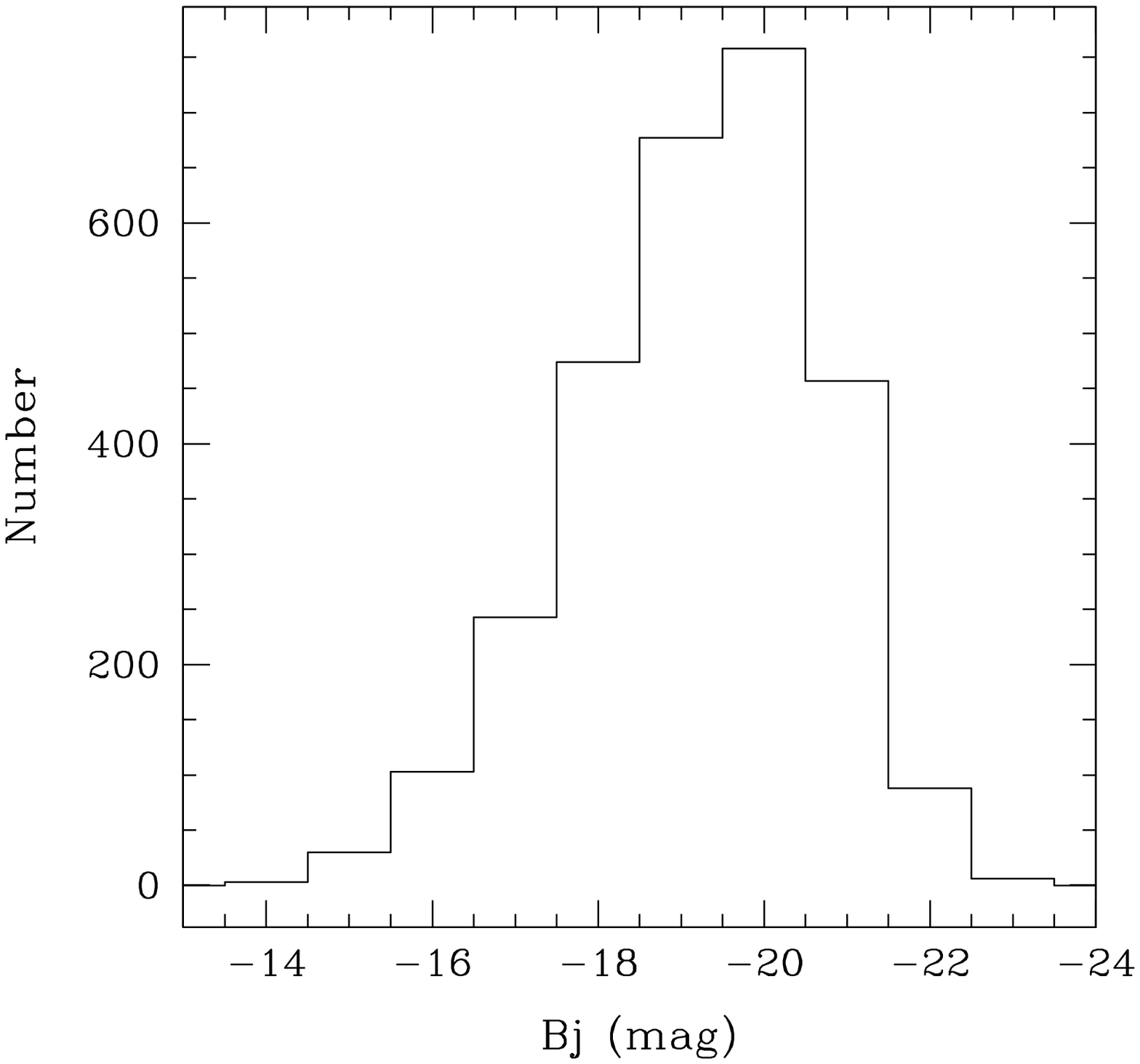}
\includegraphics[trim=-1.5cm 1cm 7cm 4cm,width=7cm,keepaspectratio=true]{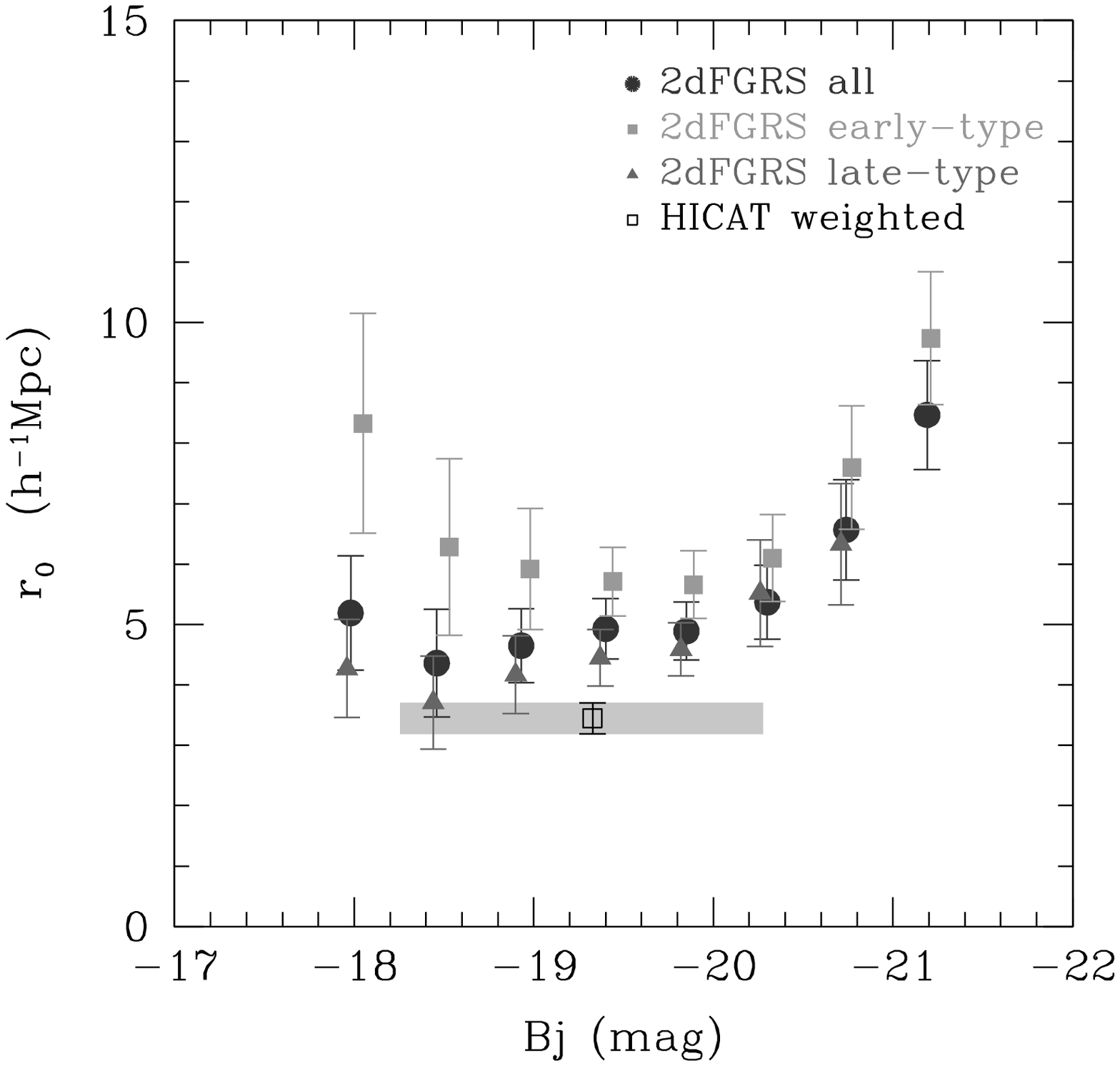}
\caption[$r_0$ luminosity dependence]{(left) $B$-band absolute magnitude distribution of galaxies in the weighted correlation function sample with \hopcat \citep{doyle2005} counterparts. (right) \hicat correlation length plotted against 2dFGRS results as a function of luminosity
\citep{norberg2002}.  The magnitude range of the grey shaded area
corresponds to the first and third quartiles of the magnitude
distribution at right and the data point corresponds to the median.
Solid squares are the 2dFGRS points for early-type galaxies, triangles
are for late-type galaxies and circlular points are the results for
all types.  All 2dFGRS points are plotted at the median of each
luminosity bin.}
\label{fig:r0}
\end{figure*}

To further compare \hicat results with those of optically-selected samples, we examine the luminosity distribution of \hicat galaxies using the \hicat optical counterpart catalogue \citep[\hopcatNS,][]{doyle2005}.  Selecting those galaxies with good optical matches and photometry, the resulting distribution for galaxies in the weighted \hicat full sample is shown in the left-hand panel of Figure~\ref{fig:r0} (74 per cent of galaxies in the sample).  The right-hand panel of Figure~\ref{fig:r0} plots the weighted full-catalogue \hicat clustering result against the luminosity dependent clustering scale lengths of the 2dFGRS \citep{norberg2002}.  From this it can be seen that \hicat galaxies span a large range of optical luminosities, and not just the lowest luminosity range most consistent with the \hicat clustering scale length.  Combined with the flatter slope of the \hicat correlation function, this indicates that \hiNS-rich galaxies exhibit weaker cluster clustering than 2dFGRS and SDSS galaxies of comparable optical luminosities, with this effect most pronounced on $<\,\, \sim1$ Mpc scales.

We calculate the correlation function for a volume limited sub-sample of \hicat as a test of the robustness of our technique.  Such a sub-sample avoids the need for any galaxy pair weighting scheme as the selection function is constant.   We did not use this sub-sample more generally as it restricts the sample size and hence accuracy with which the correlation function parameters can be determined.  However, a volume limited sample nevertheless provides an interesting check on the full catalogue results.  Applying the parameter cuts $M_{\rm HI} > 10^{9.05}\,h^{-2}M_\odot$ and $D < 30\, h^{-1}$Mpc, we retrieve a correlation function with power-law parameters $r_0 = 3.2\pm1.4$ and $\gamma=1.5\pm1.1$, in excellent agreement with those of the full sample.  

\subsubsection{Correlation Function Without Power-Law Assumption}
\label{sec:inversion}

The previous section assumes a power-law shape for the correlation function.  However, the real-space correlation function can alternatively be derived using the methods of \citet{saunders1992} and \citet{hawkins2003}, inverting the projected correlation function numerically to obtain the real space correlation function without making this assumption.  This provides an independent test of the real-space correlation function shape.  Rearranging Equation~\ref{eqn:proj_corr1} gives:

\begin{eqnarray}
\nonumber\xi(r) &=&-\frac{1}{r\pi}\frac{d}{dr}\int^{\infty}_{r}\frac{\sigma\Xi(\sigma)}{(\sigma^2-r^2)^{1/2}}d\sigma\\
&=&-\frac{1}{\pi}\int^{\infty}_{r}\frac{d\Xi(\sigma)/d\sigma}{(\sigma^2-r^2)^{1/2}}d\sigma
\end{eqnarray}

\pep Assuming $\Xi(\sigma)$ to have a step function form with values
$\Xi_i$ at logarithmic intervals with centres at $\sigma_i$, the above
integral can be evaluated ($r = \sigma_i$):

\begin{eqnarray}
\nonumber\xi(\sigma_i) = -\frac{1}{\pi}\sum_{j\geq i}\frac{\Xi(\sigma_{j+1})-\Xi(\sigma_{j})}{\sigma_{j+1}-\sigma_{j}}\times\\
\ln\left(\frac{\sigma_{j+1}+\sqrt{\sigma^2_{j+1}-\sigma^2_{i}}}{\sigma_{j}+\sqrt{\sigma^2_j-\sigma^2_i}}\right)
\end{eqnarray}

\begin{figure*}
\includegraphics[trim=0cm 1cm 2cm 0cm,width=7.2cm,keepaspectratio=true]{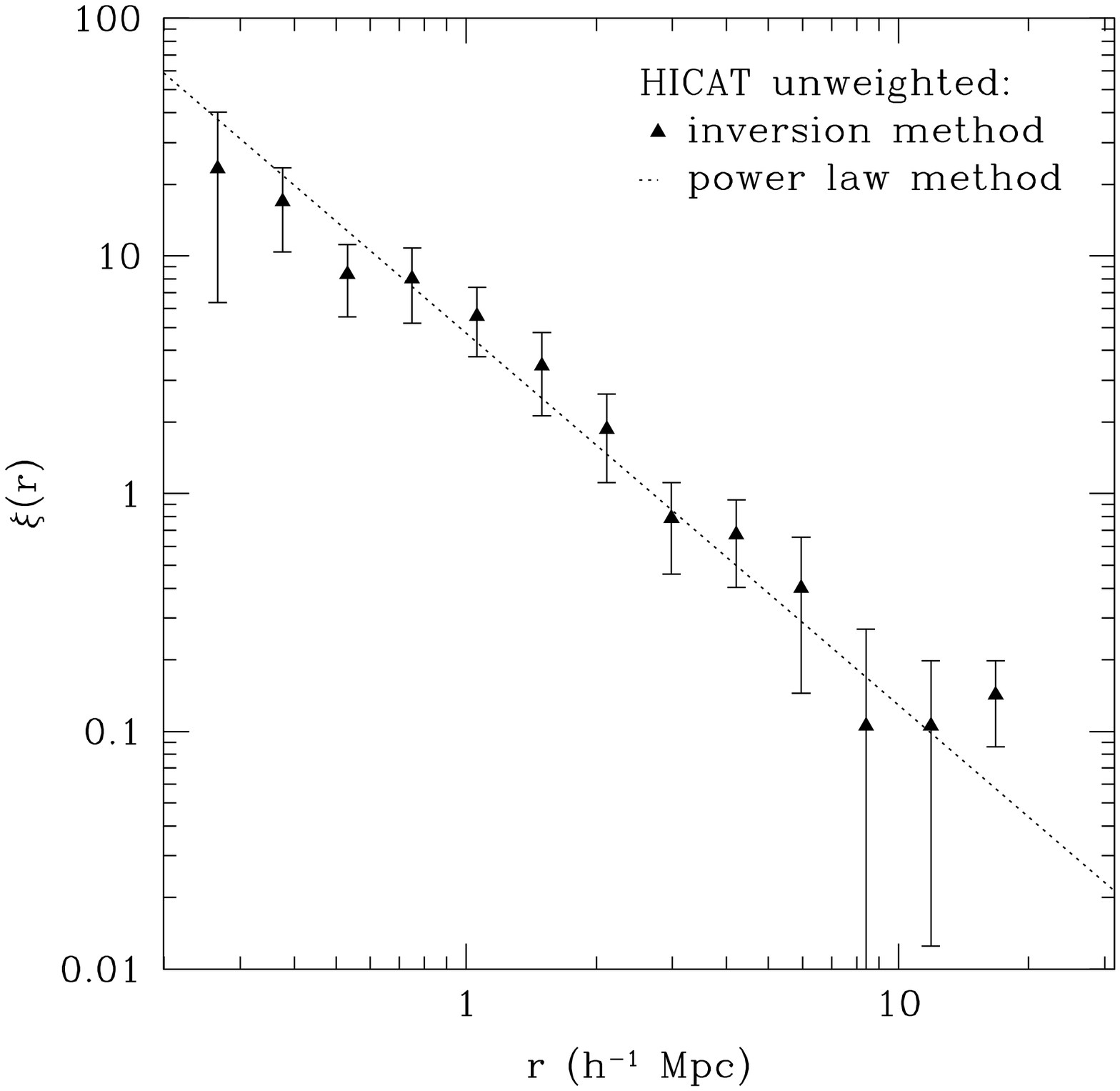}
\includegraphics[trim=-3cm 1cm 5cm 0cm,width=7.2cm,keepaspectratio=true]{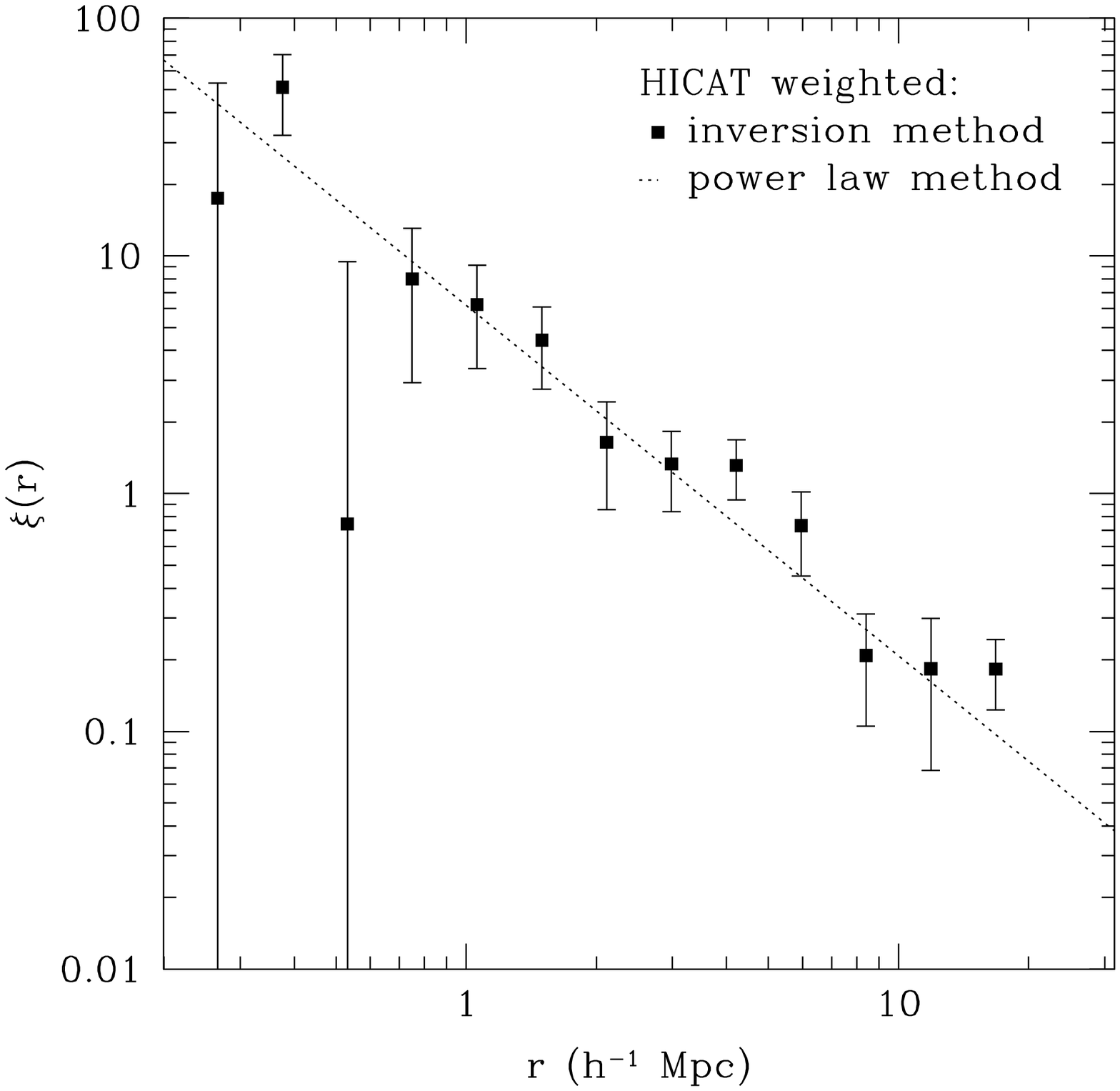}
\caption[Inversion method real-space correlation functions]{Real-space correlation function derived using inversion method (points; Section~\ref{sec:inversion}) compared to the assumed power-law real-space correlation obtained from the projected correlation function (dotted line; Section~\ref{sec:powerlaw}, fits to projected correlation function from which real-space correlation function parameters are obtained  via Equations~\ref{eqn:r0}~and~\ref{eqn:gamma} are shown in Figure~\ref{fig:cf_real_fit}): (left) unweighted, (right) weighted.}
\label{fig:cf_real_invert}
\end{figure*}

\pep The sum is truncated in the above expression (and hence the
recovered correlation function) at $30\, h^{-1}$ Mpc.  Although
exhibiting a significant amount of noise, the results of this
inversion are in excellent agreement with the correlation function
determined in the previous section assuming a power-law form
(Figure~\ref{fig:cf_real_invert}, points correspond to the inversion
method and dotted line is the result assuming a power-law).

\section{Clustering Dependencies}
\label{sec:dependencies}

From previous studies \citep{norberg2001,norberg2002, zehavi2005} it has been found that galaxy clustering varies as a function of luminosity, with the most luminous galaxies showing the strongest clustering.  A great strength of using an \hiNS-selected catalogue is that it uniquely provides the additional ability to study clustering as a function of gas content (via \hi mass) and halo mass (via rotational velocity).  We use optical data from \hopcat to compare these dependencies with the observed luminosity trend.  Results are summarised in Table~\ref{tab:r0gamma_mass} with a more detailed description for each of the parameters given below.

\subsection{Luminosity}

\begin{figure*}
\begin{minipage}{\minipagesize}
\begin{center}
\includegraphics[trim=2cm 1cm 2cm 1cm,width=7cm,keepaspectratio=true]{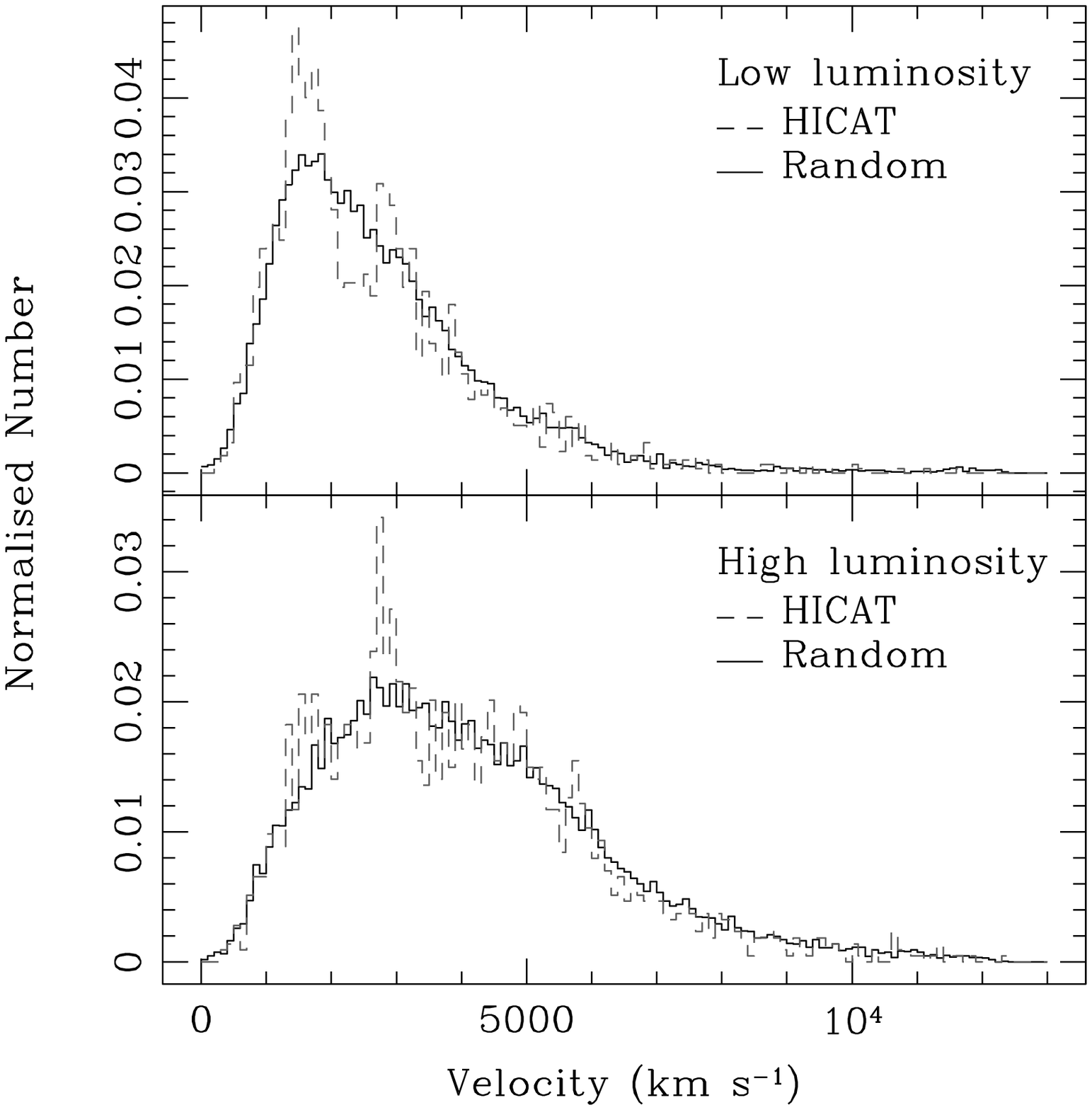}
\caption{Low (top) and high (bottom) luminosity sample radial velocity distributions. Threshold luminosity is Bj = -19.5.  Dashed line shows the distribution from \hicat and the solid line that of the
random sample.}
\label{fig:cf_lum_rhist}
\end{center}
\end{minipage}
\end{figure*}

\begin{figure*}
\begin{center}
\hspace{-11mm}
\includegraphics[trim=1cm 0cm 0.6cm -1cm,width=7cm,keepaspectratio=true]{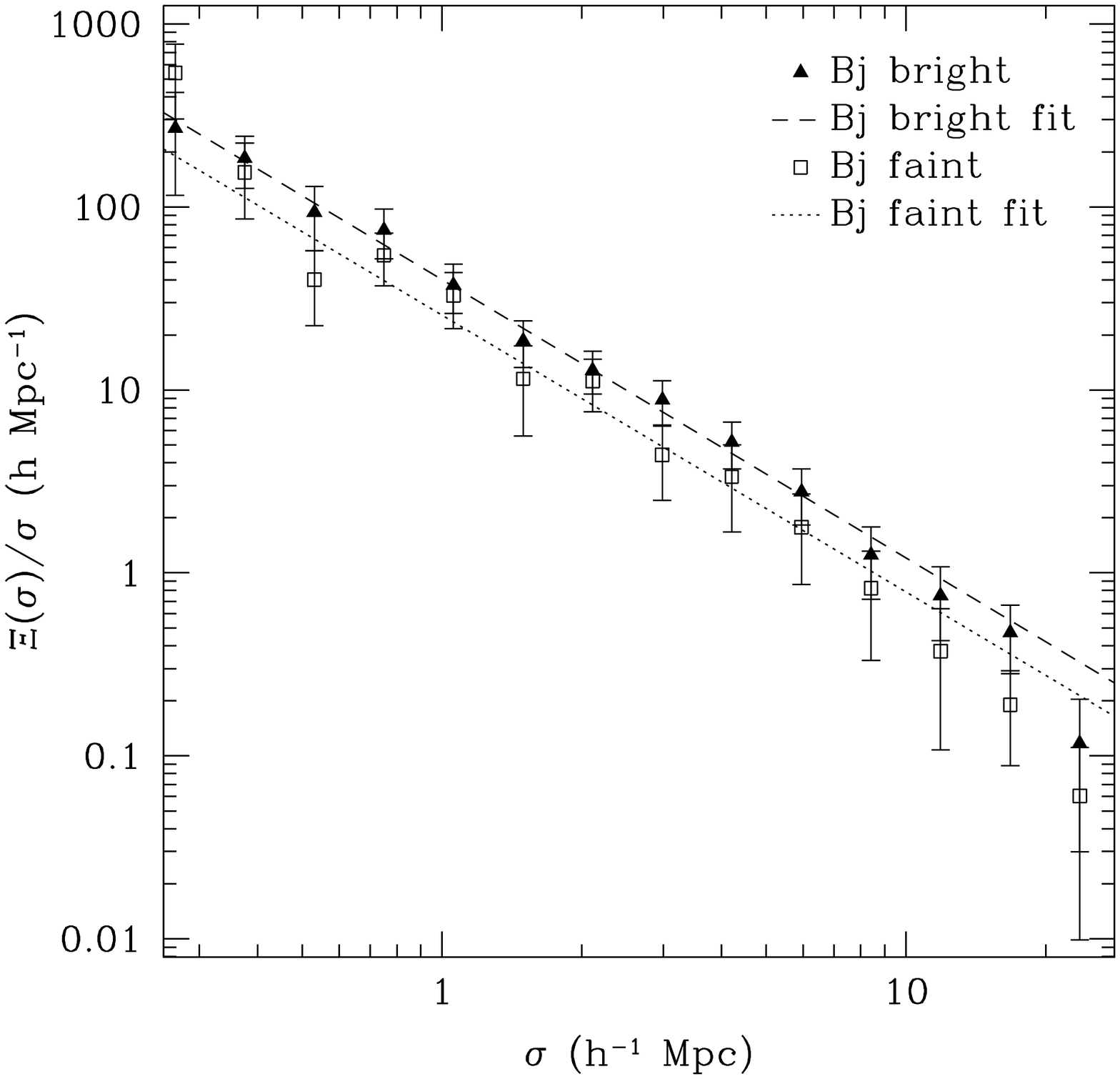}
\hspace{6mm}
\includegraphics[trim=1.0cm 1cm 3.5cm 2cm,width=7cm,keepaspectratio=true]{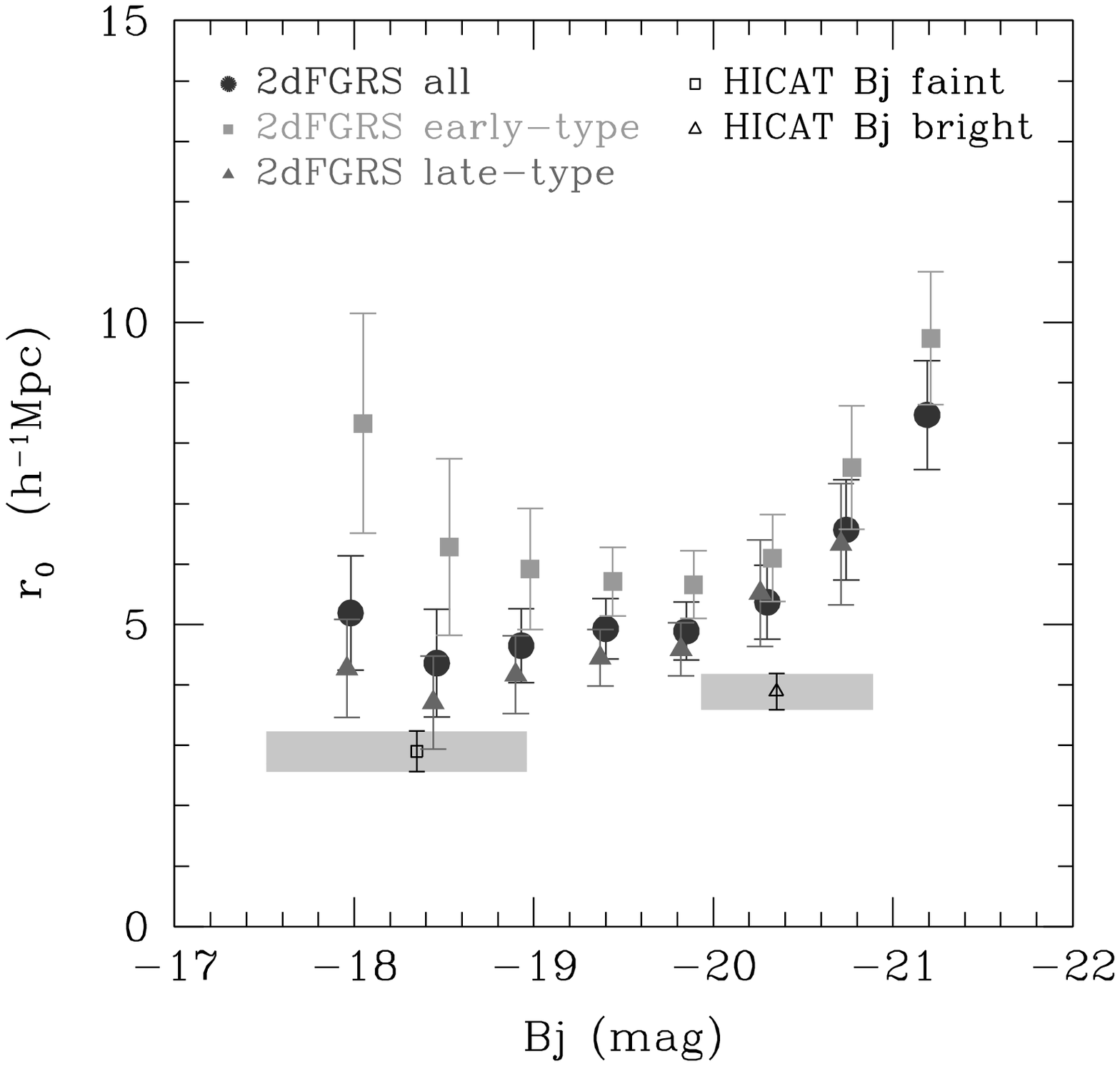}
\caption[]{(left) Weighted projected real-space correlation functions with power-law fits. Fitting restricted to points $\sigma<10\, h^{-1}$Mpc. Threshold luminosity is Bj = -19.5. (right) Low and high luminosity correlation lengths plotted against 2dFGRS results as a function of luminosity \citep{norberg2002}.  The magnitude range of the grey shaded area corresponds to the first and third quartiles of the \hopcat \citep{doyle2005} magnitude distribution  and the data point corresponds to the median. Solid squares are the 2dFGRS points for early-type galaxies, triangles are for late-type galaxies and circlular points are the results for all types.  All 2dFGRS points are plotted at the median of each luminosity bin.}
\label{fig:cf_lum_fit}
\end{center}
\end{figure*}

The dependence of galaxy clustering is examined by dividing the sample in two around a luminosity of Bj = -19.5.  As before, magnitudes are calculated using measurements from \hopcatNS.  Galaxies not matched or having data of insufficient quality are given random magnitudes generated from the observed luminosity distribution (26 per cent of galaxies in the weighted sample).  Random magnitudes are used rather than those from an estimation method, such as deriving luminosities from the observed \hi masses, as this could make it difficult to disentangle the different clustering dependencies. These galaxies will dilute the observed clustering dependence, but are maintained in the calculation to ensure the on-sky consistency between the real and random catalogue.  The radial velocity distribution histograms for the sub-samples and their generate random catalogues are shown in Figure~\ref{fig:cf_lum_rhist}.  The left-hand panel of Figure~\ref{fig:cf_lum_fit} shows the fitted projected correlation function yielding clustering parameters of  $r_0 = 2.90 \pm 0.33$ with $\gamma=1.51\pm0.14$ for the low luminosity sample, and $r_0 = 3.89 \pm 0.30$ with $\gamma=1.52\pm0.10$ for those with higher luminosities.   The right-hand panel of Figure~\ref{fig:cf_lum_fit} plots the calculated clustering scale lengths against the 2dFGRS results.  Grey shaded areas correspond to the first and third luminosity quartiles for each sample.  From this it can be seen that although overall more weakly clustered, the \hicat galaxies exhibit a luminosity clustering dependence consistent with that observed for optically selected galaxies \citep{norberg2001,norberg2002, zehavi2005}.  Correlation function slopes for the two sub-samples are nearly identical. 

\subsection{\hi Mass}

\begin{figure*}
\begin{minipage}{\minipagesize}
\begin{center}
\includegraphics[trim=2cm 1cm 2cm 1cm,width=7cm,keepaspectratio=true]{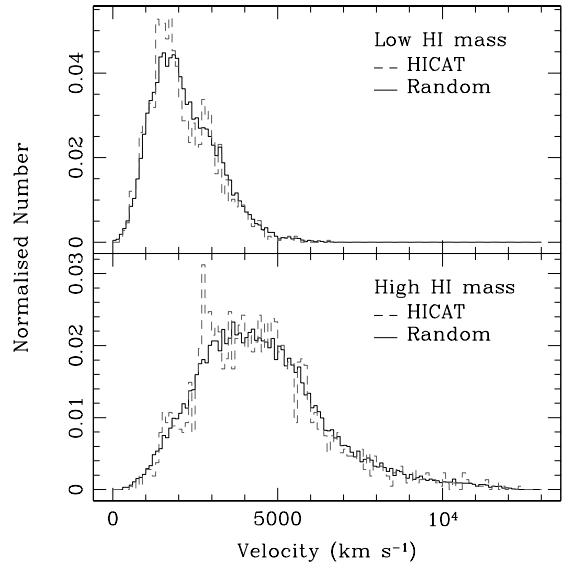}
\caption{Low (top) and high (bottom) \hi mass sample radial velocity distributions.  Threshold \hi mass is $10^{9.25}\,h^{-2}M_{\odot}$.  Dashed line shows the distribution from \hicat and the solid line that of the random sample.}
\label{fig:cf_mass_rhist}
\end{center}
\end{minipage}
\end{figure*}

\begin{figure*}
\begin{center}
\hspace{-11mm}
\includegraphics[trim=1cm 0cm 0.6cm -1cm,width=7cm,keepaspectratio=true]{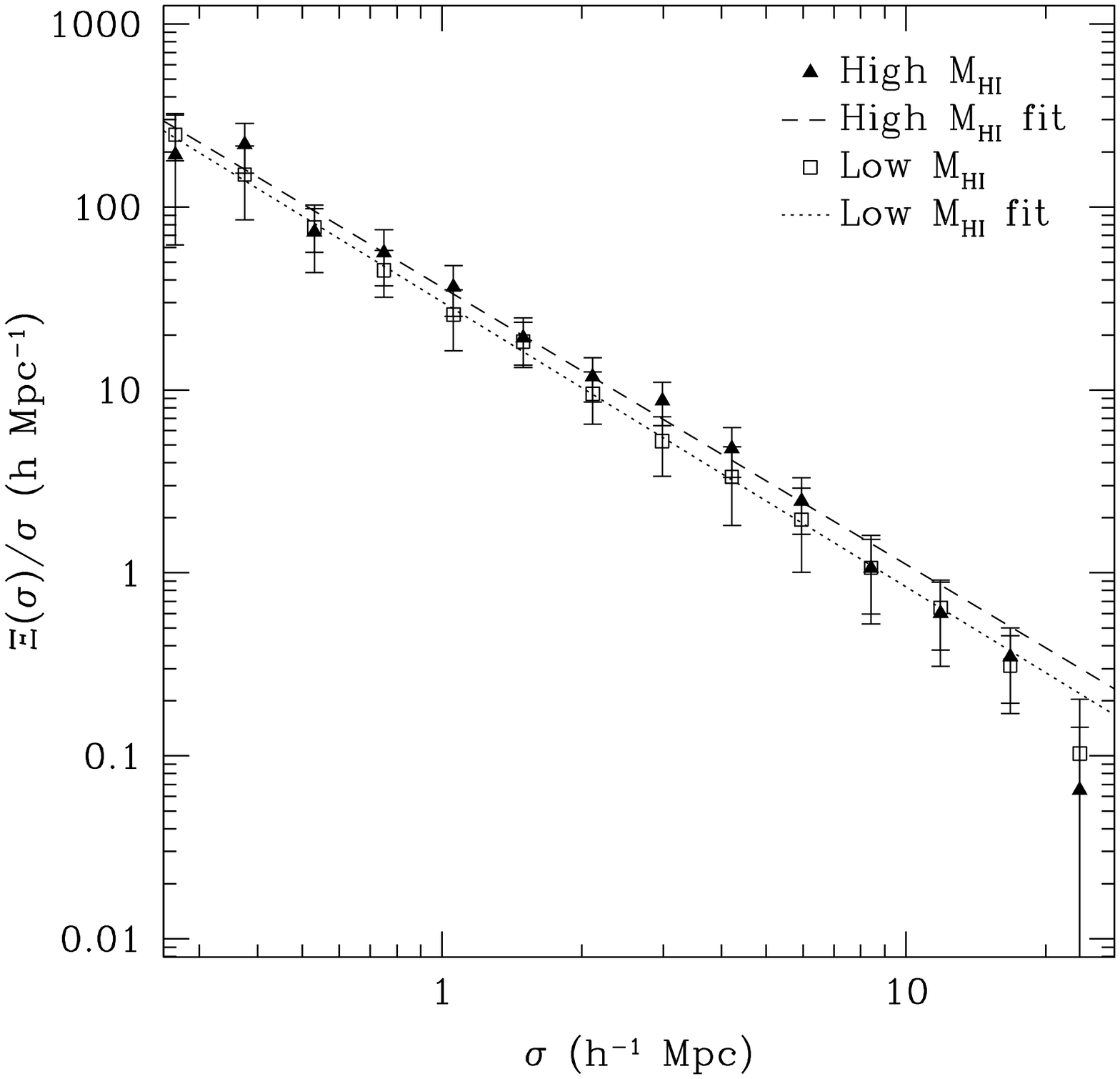}
\hspace{6mm}
\includegraphics[trim=1.0cm 1cm 3.5cm 2cm,width=7cm,keepaspectratio=true]{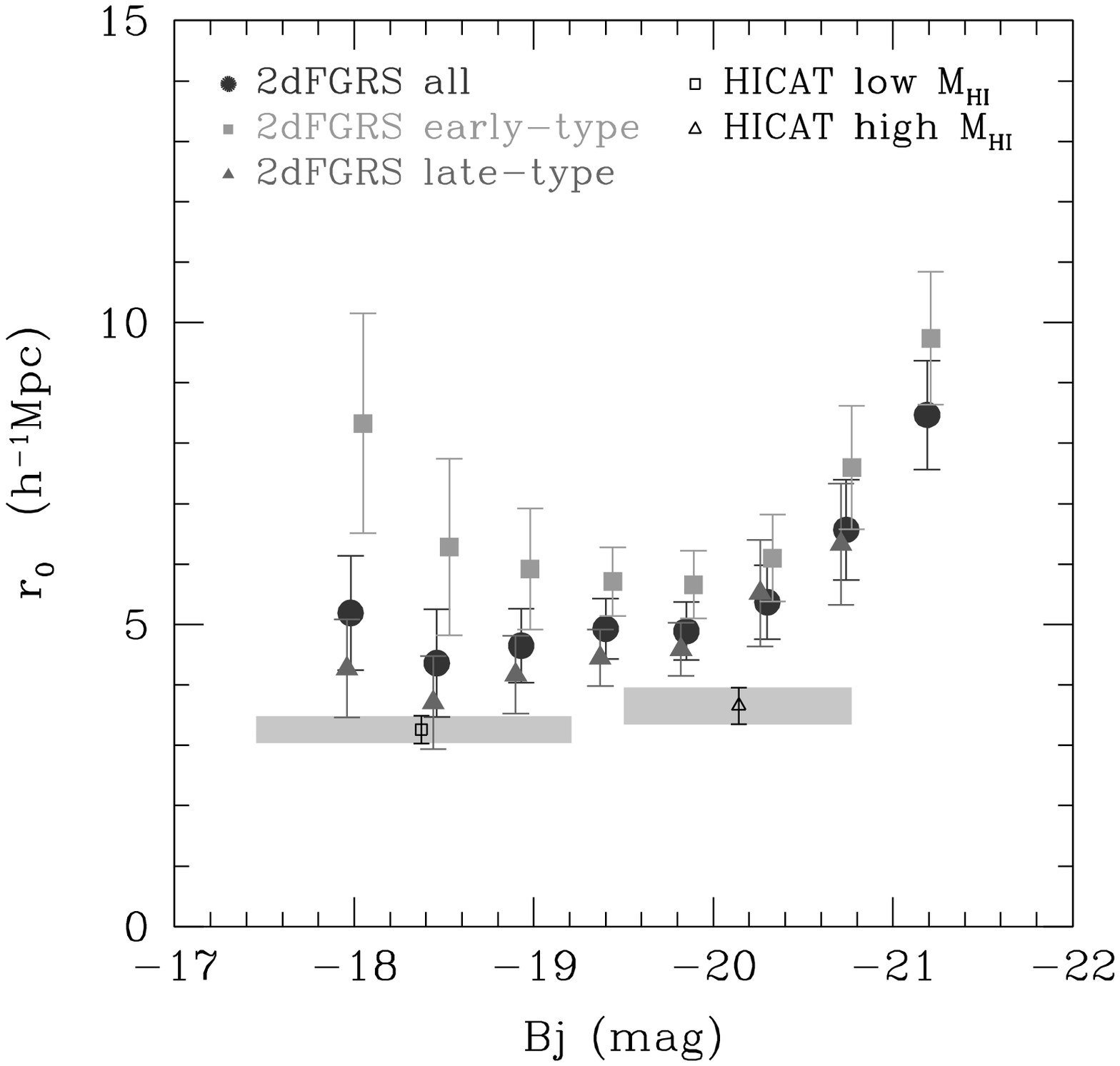}
\caption[]{(left) Weighted projected real-space correlation functions with power-law fits. Fitting restricted to points $\sigma<10\, h^{-1}$Mpc.  Threshold mass is $\mhi = 10^{9.25}\,h^{-2}M_{\odot}$. (right) Low and high \hi mass correlation lengths plotted against 2dFGRS results as a function of luminosity \citep{norberg2002}.  The magnitude range of the grey shaded area corresponds to the first and third quartiles of the \hopcat \citep{doyle2005} magnitude distribution and the data point corresponds to the median. Solid squares are the 2dFGRS points for early-type galaxies, triangles are for late-type galaxies and circlular points are the results for all types.  All 2dFGRS points are plotted at the median of each luminosity bin.}
\label{fig:cf_mass_fit}
\end{center}
\end{figure*}

We now divide the sample in two around an \hi mass of $10^{9.25}\,h^{-2}M_{\odot}$.  This corresponds to a mass $\sim M_{\rm HI}^*/4$. Figure~\ref{fig:cf_mass_rhist} plots the radial velocity distributions for these two samples and their corresponding random catalogues. The projected real-space correlation functions with power-law fits are shown in the left-hand panel of Figure~\ref{fig:cf_mass_fit}. The right-hand panel of Figure~\ref{fig:cf_mass_fit} compares the \hi mass dependent \hicat scale lengths with the 2dFGRS luminosity dependent values of \citet{norberg2002}.  From the weighted results, clustering of the low \hi mass galaxies ($r_0 = 3.26 \pm 0.23$, $\gamma=1.56\pm0.11$) is only marginally lower than that of high mass galaxies ($r_0 = 3.65 \pm 0.30$, $\gamma=1.51\pm0.10$).  At the mass limits examined, \hi mass does not therefore provide a robust method for selecting the most strongly clustered objects, as can be done with stellar luminosity.  This is consistent with the relative depletion of \hi mass relative to stellar luminosity in more strongly clustered environments.

\subsection{Rotational Velocity}

An alternative parameter available in \hicat to test for clustering dependence is rotational velocity.  This is interesting as rotational velocity is the observable quantity most directly linked to the total halo mass.   As such, the dependence of galaxy clustering on halo mass can be tested and compared to simulations, without having to relate halo properties to alternative observables such as optical luminosity which involve more complicated and poorly understood physics.  \hicat offers a unique ability to assess this dependence through the availability of 21cm linewidths for all galaxies in the sample.  These are converted to rotational velocities by applying a simple correction for inclination ($w_{50}$ is the width at 50 per cent of the 21cm profile peak flux):

\begin{equation}
v_{\rm rot} = \frac{w_{\rm 50}}{2 \sin(i)}
\end{equation}

\pep Inclinations $i$ are determined from the observed \hopcat axial ratio where a good optical counterpart idenification is available:

\begin{equation}
i = \arccos \left( \sqrt{ \frac{(b/a)^2 - (b/a)_{\rm eos}^2}{1-(b/a)_{\rm eos}^2} } \right)
\end{equation}
 
\pep Here, $(b/a)$ is the semi-minor to semi-major axis ratio, and $(b/a)_{\rm eos}$ is axial ratio for an edge-on spiral galaxy, and is set to 0.1.  Galaxies without good optical identifications are given random inclinations (26 per cent of the sample).  

\begin{figure*}
\begin{minipage}{\minipagesize}
\begin{center}
\includegraphics[trim=2cm 1cm 2cm 1cm,width=7cm,keepaspectratio=true]{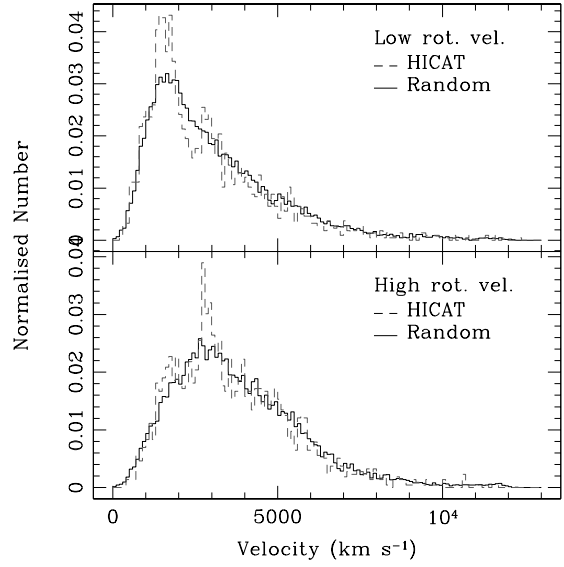}
\caption{Low (top) and high (bottom) rotational velocity sample radial velocity distributions.  Threshold rotational velocity is 108 \kmsNS.  Dashed line shows the distribution from \hicat and the solid line that of the random sample.}
\label{fig:cf_rotvel_rhist}
\end{center}
\end{minipage}
\end{figure*}

\begin{figure*}
\begin{center}
\hspace{-11mm}
\includegraphics[trim=1cm 0cm 0.6cm -1cm,width=7cm,keepaspectratio=true]{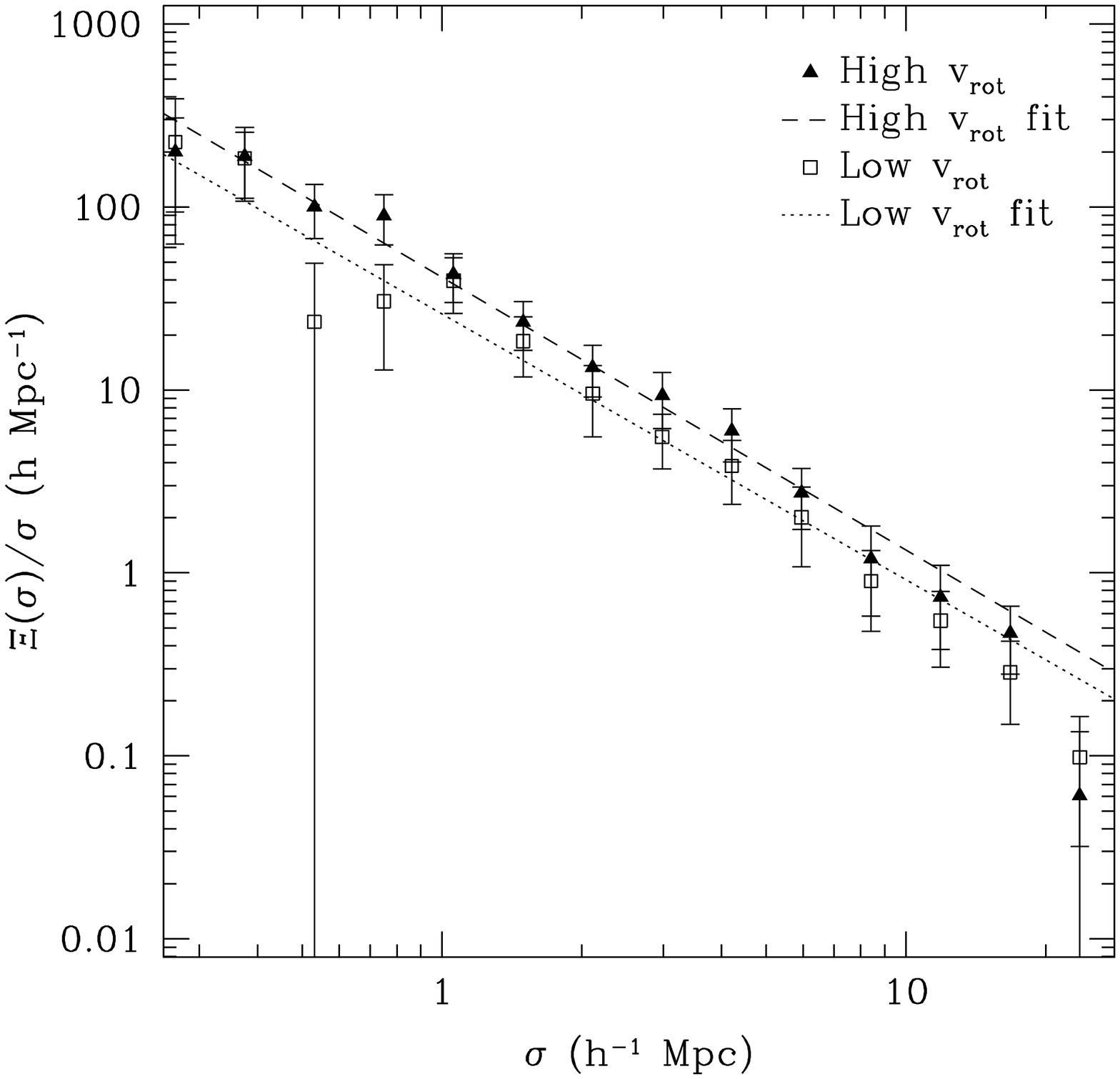}
\hspace{6mm}
\includegraphics[trim=1.0cm 1cm 3.5cm 2cm,width=7cm,keepaspectratio=true]{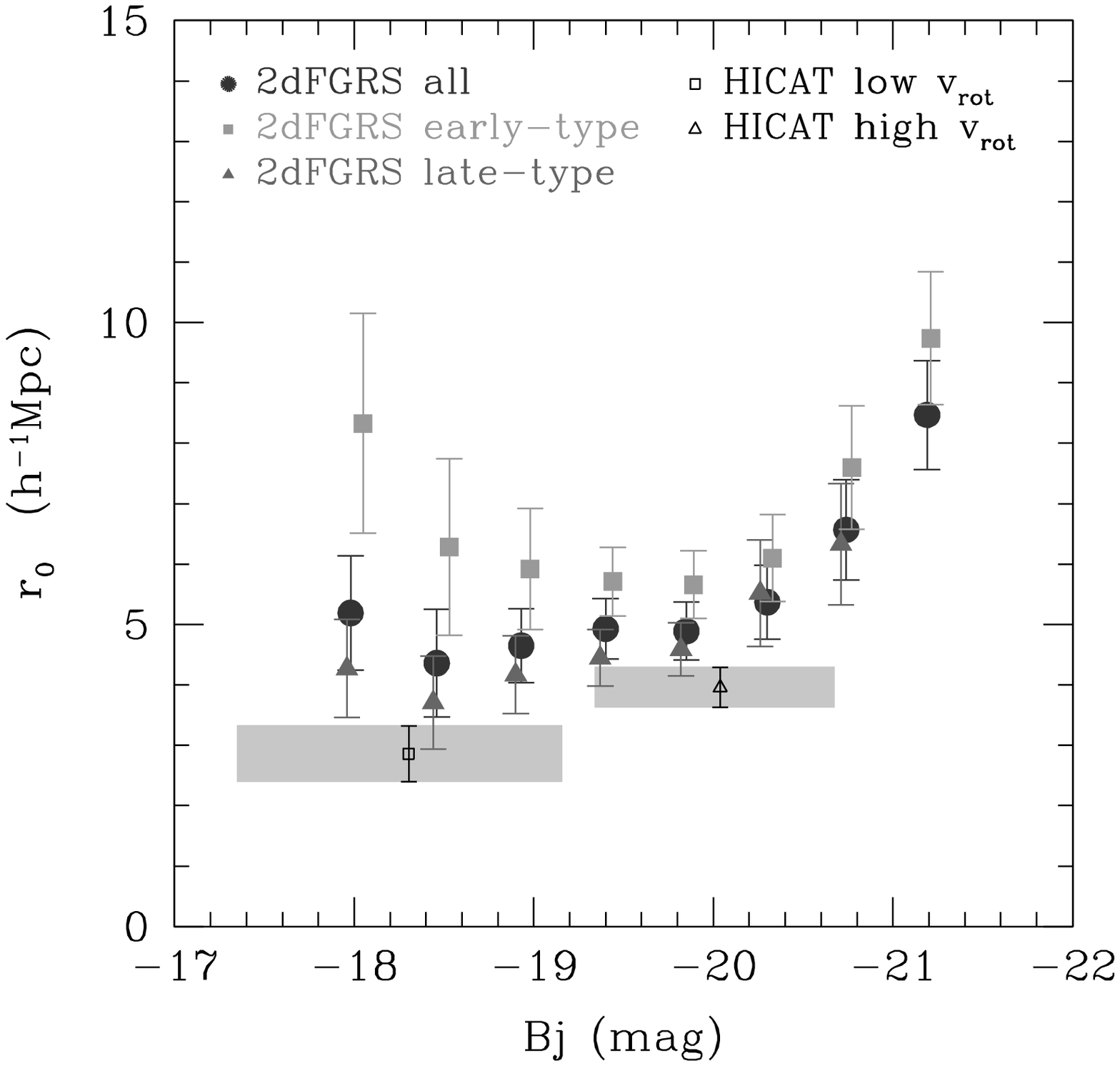}
\caption[]{(left) Weighted projected real-space correlation functions with power-law fits.  Fitting restricted to points $\sigma<10\, h^{-1}$Mpc.  Threshold rotational velocity is 108 \kmsNS. (right) Low and high rotational velocity correlation lengths plotted against 2dFGRS results as a function of luminosity \citep{norberg2002}.  The magnitude range of the grey shaded area corresponds to the first and third quartiles of the \hopcat \citep{doyle2005} magnitude distribution and the data point corresponds to the median. Solid squares are the 2dFGRS points for early-type galaxies, triangles are for late-type galaxies and circular points are the results for all types.  All 2dFGRS points are plotted at the median of each luminosity bin.}
\label{fig:cf_rotvel_fit}
\end{center}
\end{figure*}

The threshold rotational velocity used to divide the sample is 108 \kmsNS, which we roughly estimate corresponds to a sub-halo mass of $\sim 10^{11}M_\odot$ \citep{bullock2001}.   The radial velocity distributions of each sub-sample and their corresponding random samples are shown in Figure~\ref{fig:cf_rotvel_rhist}.  The left-hand panel of Figure~\ref{fig:cf_rotvel_fit} plots the projected real-space correlation functions, with the fits yielding real-space power-law parameters of $r_0 = 2.86 \pm 0.46$ with $\gamma=1.45\pm 0.14$ for the low rotational velocity sample, and $r_0 = 3.96 \pm 0.33$ with $\gamma=1.49\pm0.10$ for galaxies with high rotational velocities.   As both stellar mass and rotational velocity are correlated with dark matter halo mass, it is not unexpected that spatial clustering increases with both stellar mass and rotational velocity in our sample. The comparison of these results to the 2dF luminosity dependent values is given in the right-hand panel of Figure~\ref{fig:cf_rotvel_fit}. 

\section{Comparison with CDM simulations}
\label{sec:Millennium}

Errors in the measured values of $r_0$ and $\gamma$ are all computed using the jackknife method. One concern is that many of the large scale structures seen in \hipass are larger than the total \hipass survey volume. Therefore, different jackknife sub-samples are not independent, which might result in an underestimation of the uncertainties. In order to test this effect, we make use of the Millennium Run \citep{springel2005} to construct independent synthetic \hipass volumes. 

\citet{croton2006} use a semi-analytical prescription to assign cold gas masses to individual dark matter halos identified in the Millennium Run. This prescription includes detailed modelling of cooling, star formation, supernova feedback, galaxy mergers and metal enrichment. The cold gas masses include both \hiNS, molecular hydrogen (H$_2$) and He. We take the ratio of \hi mass to total cold gas mass to be 0.5, which is roughly derived by assuming molecular gas masses to be $\sim 50$ per cent of the \hi gas mass \citep[note that considerable variation is observed in the M(H$_2$)/M(\hiNS) ratio,][]{young1989}, and a 25 per cent mass fraction of helium.  Within the total 500$^3$ $h^{-3}$ Mpc$^3$ box we identify 16 independent volumes of  120$^3$ $h^{-3}$ Mpc$^3$. From each of these boxes we select synthetic \hicat samples by placing an `observer' on the edge of the box and then using the selection function described in \citet{zwaan2004} to select galaxies. We choose to only select galaxies with \hi masses larger than $M_{\rm HI}=10^{8.7}\,h^{-2}M_{\odot}$, roughly corresponding to the mass limit of the Millennium Run.  We also constructed two further sets of synthetic samples corresponding to the high \hi mass ($M_{\rm HI} \ge 10^{9.25}\,h^{-2}M_{\odot}$) and high luminosity ($Bj \le -19.5 $) sub-samples.  Synthetic catalogues are not made for the other sub-samples as the corresponding low \hi mass and faint galaxy samples are not well represented in the Millennium Run data, and $v_{rot}$ is not directly available in the \citet{croton2006} database.

As a first test we use the bivariate stepwise maximum likelihood method from \citet{zwaan2005} to construct \hi mass functions from the full synthetic samples. We find that the \hi mass functions are in excellent agreement with the real \hipass  \hi mass function, providing confidence in the generated samples.  \citealt{croton2006} also find that their semi-analytical results can accurately reproduce the field optical galaxy luminosity function. 

The projected correlation function is calculated for each sample and negligible 0.001 errors ascribed to each datapoint for the power-law fitting.  A separation limit of $\sigma<10\, h^{-1}$Mpc is again applied for the fitting, as was done for the real dataset.  From this we find the mean and standard deviation of the derived real-space power-law correlation function parameters to be $r_0 = 3.49 \pm 0.43$ and $\gamma  = 1.35 \pm 0.12$ for the full dataset. These are in good agreement with the results from \hicatNS.  For the high luminosity samples we find that $r_0 = 3.69 \pm 0.41$ and $\gamma = 1.39 \pm 0.09$, and for the high \hi mass samples we find that $r_0 = 3.83 \pm 0.35$ and $\gamma = 1.40 \pm 0.07$.  The larger errors on the correlation function length $r_0$ (by 72 \% in the case of the full sample, 37 \% for the high luminosity sample, and 17 \% for the high \hi mass sample) indicate that we may slightly underestimate the \hicat correlation function errors from jackknife as a result of the relatively small \hipass volume, while the errors on $\gamma$ are generally consistent.  However, even taking these larger error values for $r_0$, our principal conclusions remain unchanged.

\section{Discussion}
\label{sec:discussion}

Our results indicate that \hiNS-rich galaxies are among the most weakly clustered objects known.   That the clustering of \hiNS-rich objects is weak is also in first order agreement with the weaker clustering of galaxies with active star formation \citep{madgwick2002}, faint \citep{zehavi2005} and late-type \citep{norberg2002} galaxies.  It is also consistent with the weaker clustering of \hii galaxies, which are usually gas-rich dwarf systems \citep[$r_0 = 2.7 h^{-1}$ Mpc, 425 galaxies;][]{iovino1988}.  There are a number of effects that could contribute to the lower observed clustering of \hiNS-rich galaxies compared to the optically selected galaxy population.  

One important factor is the effect of environment on \hi gas content.  It has already been well established that there are few \hiNS-rich galaxies near the cores of rich clusters \citep[e.g.][]{waugh2002}. Possible processes that can remove \hi from galaxies from galaxies in the densest environments include: the stripping of \hi gas by tidal effects in galaxy concentrations \citep[e.g., galaxy harassment,][or by the overall concentration potential]{moore1996}, ram pressure stripping \citep{gunn1972} or strangulation \citep{balogh2000}.  An increased rate of tidal interactions may also trigger increased star-formation \citep{barton2000}, which in turn depletes galaxy \hi gas content.  It is also worth noting that galaxy properties have been observed to vary at substantial distances from the centers of clusters \citep{lewis2002,gomez2003,balogh2004,zwaan2005}.  

If environmental effects leading to a depletion of \hi gas are responsible, it might be expected that there should exist a strongly clustered counterpart population corresponding to those galaxies which have had their \hi gas removed.  While no direct evolutionary links can be drawn, there do exist galaxy populations which could meet this critera, such as the $L<L_*/3$ red galaxies identified in \citet{hogg2003}.  Interestingly, \citet{norberg2002} also identified stronger clustering for low luminosity early-type galaxies, although the result was not viewed as significant.  These faint red galaxies represent a strongly clustered low mass galaxy population with little or no star formation, and that preferentially reside in the very massive dark matter halos of clusters. 

Another possibility for the lower observed clustering of \hiNS-rich galaxies is that they form in different, intrinsically less clustered, dark matter halos compared to galaxies selected in the optical.  As noted by \citet{norberg2002}, the luminosity dependence of clustering is consistent with the results of CDM simulations: the brightest galaxies form in the most clustered and massive dark matter halos \cite[e.g.,][]{benson2001}.  Similarly, the dependence of clustering on morphology may also reflect a relation between the morphology of a galaxy and the parameters of its halo.  In this vein, the lower clustering of \hiNS-rich galaxies may therefore be the result of \hiNS-rich galaxies only forming in the low-medium peaks of the initial density field that are yet to have been accreted onto the most massive and strongly clustered halos.  This possibility has been suggested for the lower clustering observed for low surface brightness galaxies \citep{mo1994}.  Recent results by \citet{gao2005} also indicate that dark halo clustering is a strong function of halo age, with the youngest halos being the most weakly clustered.  Moreover, this dependence is found to increase with decreasing mass.  As such, if \hiNS-rich galaxies preferentially form in low mass halos, any tendency toward younger halos would act to further decrease the strength of \hiNS-rich galaxy clustering.  

Our observations that \hiNS-rich galaxies are particularly weakly clustered, and that the clustering strength of galaxies depends on rotational velocity (and by implication halo mass) are consistent with the biasing of \hiNS-rich galaxies toward low mass halos.  Both environmental factors and initial conditions may contribute to this result.  Also, the similar clustering dependence of \hiNS-rich galaxies on stellar mass and rotational velocity compared to the weaker dependence on \hi mass argues for stellar mass being a better tracer of halo mass than \hi gas mass.

\section{Conclusions}
\label{sec:conclusions}

Existing studies of galaxy clustering find strong dependencies on a number of parameters, highlighting an underlying trend for clustering to be strongest for more luminous galaxies \citep{norberg2001,norberg2002}, earlier morphological types \citep[e.g.][]{loveday1995}, and galaxies with old stellar populations \citep{norberg2002,zehavi2002,madgwick2003}.  

The low clustering measured for \hiNS-rich galaxies is consistent with these trends, \hiNS-rich galaxies having preferentially spiral morphologies, active star formation and blue colors.  However, the scale length of $r_0 = 3.45 \pm 0.25\,h^{-1}$ Mpc and slope $\gamma=1.47 \pm 0.08$ place \hiNS-rich galaxies among the most weakly clustered objects known and at the extreme weak end of the observed clustering distribution.  Compared to results from the 2dFGRS, \hiNS-rich galaxies are also more weakly clustered than optically selected galaxies of similar luminosities.

Dividing the \hicat sample by \hi mass around a threshold of \mhi = $10^{9.25}\,h^{-2}M_{\odot}$, only a very marginal dependence of galaxy clustering strength on \hi mass is observed.   The scale length for the low \hi mass sample is found to be $r_0 = 3.26\pm0.23\,h^{-1}$Mpc and for the high mass sample $r_0 = 3.65\pm0.30\,h^{-1}$Mpc.   Alternatively dividing the sample on the basis of rotational velocity, a stronger dependence is seen.  The clustering scale length for galaxies $v_{\rm rot} < 108$ \kms is $r_0 = 2.86 \pm 0.46\,h^{-1}$ Mpc compared to $r_0 = 3.96 \pm 0.33\,h^{-1}$ Mpc for the high rotational velocity sample.   This is similar to the luminosity trend, where $r_0 = 2.90 \pm 0.33$ for galaxies $Bj > -19.5$ and $r_0 = 3.89 \pm 0.30$ for $Bj \le -19.5$.  

Our results are consistent with galaxy clustering being fundamentally a function of halo mass, which is well traced by stellar luminosity but poorly traced by HI gas mass.  In this scenario, \hiNS-rich galaxies preferentially occupy lower mass halos compared to the general galaxy population, accounting for their low clustering strength.  Both environmental processes and initial conditions may lead to this effect.

\acknowledgements

The \hipass and \hicat teams are acknowledged their role in planning and executing the programs which created the datasets from which this work is derived.  We also thank Peder Norberg for providing us with the 2dF correlation function results for comparison and Darren Croton for his helpful advice on the Millennium Run data.  The Millennium Run simulation used in this paper was carried out by the Virgo Supercomputing Consortium at the Computing Centre of the Max-Planck Society in Garching. The semi-analytic galaxy catalogue is publicly available at http://www.mpa-garching.mpg.de/galform/agnpaper.


\bibliographystyle{apj}
\bibliography{ms}


\clearpage
\begin{table*}
\small
\begin{center}
\begin{tabular}{p{55mm}p{55mm}p{20mm}p{20mm}}
\hline
\hline
Sample & Selection Criteria & $r_0$         & $\gamma$\\ 
       &                    & ($h^{-1}$Mpc) & \\
\hline                                                                
\hicat (weighted)$^1$        & $-$                                                  & $3.45\pm0.25$    & $1.47\pm0.08$\\
2dFGRS Late-Type Faint$^2$   & $-19.0 < M_{b_J}-5\log_{10}h<-18.0$   & $3.71\pm0.77$  & $1.76\pm0.11$\\
2dFGRS Late-Type Bright$^2$  & $-21.5 < M_{b_J}-5\log_{10}h<-20.5$   & $6.33\pm1.01$  & $2.01\pm0.29$\\
2dFGRS Early-Type Faint$^2$  & $-20.5 < M_{b_J}-5\log_{10}h<-19.5$   & $5.66\pm0.56$  & $1.87\pm0.09$\\
2dFGRS Early-Type Bright$^2$ & $-22.0 < M_{b_J}-5\log_{10}h<-21.0$   & $9.74\pm1.16$  & $1.95\pm0.37$\\
SDSS Faint$^3$                            & $-18 < M_r < -17$                                      & $2.68\pm0.39$ & $1.99\pm0.09$\\
SDSS Bright$^3$                           & $-23 < M_r < -22$                                      & $10.04\pm0.37$ &$2.04\pm0.08$\\     
\hline
\hline
\multicolumn{4}{l}{$^1$This work, $^2$\cite{norberg2002}, $^3$\cite{zehavi2005}}\\
\end{tabular}
\caption[$r_0$ and $\gamma$ values]{Measured values of $r_0$ and
$\gamma$ for \hicatNS, 2dFGRS and SDSS galaxies}
\label{tab:r0gamma}
\end{center}
\end{table*}

\begin{table*}
\small
\begin{center}
\begin{tabular}{p{50mm}p{35mm}p{18mm}p{18mm}p{18mm}}
\hline
\hline
Sample                               & Selection Criteria                                               & $r_0$               & $\gamma$   & Number of\\ 
                                             &                                                                               & ($h^{-1}$Mpc)&                        & Galaxies              \\
\hline                                                                
Low \hi Mass                            & $M_{\rm HI} < 10^{9.25}h^{-2}M_{\odot}$     & $ 2.63\pm 0.23$   & $1.59\pm 0.12$  &  2094\\
Low \hi Mass (weighted)        & $M_{\rm HI} < 10^{9.25}h^{-2}M_{\odot}$     & $ 3.26\pm 0.23$   & $1.56\pm 0.11$  &  2093\\
High \hi Mass                           & $M_{\rm HI} \ge 10^{9.25}h^{-2}M_{\odot}$  & $ 3.23\pm 0.27$   & $1.45\pm 0.09$  &  2082\\
High \hi Mass (weighted)       & $M_{\rm HI} \ge 10^{9.25}h^{-2}M_{\odot}$  & $ 3.65\pm 0.30$   & $1.51\pm 0.10$  &  1727\\
Low Luminosity                        & $Bj > -19.5$                                                         &$ 2.51 \pm 0.23$   & $1.61\pm 0.13$  &  2097\\
Low Luminosity (weighted)    & $Bj > -19.5$                                                         &$ 2.90 \pm 0.33$   & $ 1.51\pm 0.14$ & 2024\\
High Luminosity                       & $Bj \le -19.5 $                                                      &$ 3.21 \pm 0.21$   & $ 1.53\pm 0.09$  & 2079\\
High Luminosity (weighted)   & $Bj \le -19.5 $                                                      &$ 3.89 \pm 0.30$   & $ 1.52 \pm 0.10$ & 1796 \\
Low $v_{\rm rot}$                     & $v_{\rm rot} < 108$ \kms                                   &$ 2.50\pm 0.24$   & $1.61\pm 0.13$   & 2093\\
Low $v_{\rm rot}$ (weighted) & $v_{\rm rot} < 108$ \kms                                   &$ 2.86\pm 0.46$   & $1.45\pm 0.14$   & 1953\\
High $v_{\rm rot}$                    & $v_{\rm rot} \ge 108$ \kms                                &$ 3.11\pm 0.20$   & $1.56\pm 0.10$   & 2089\\
High $v_{\rm rot}$ (weighted)& $v_{\rm rot} \ge 108$ \kms                                &$ 3.96\pm 0.33$   & $1.49\pm0.10$    & 1877\\
All                                               & $-$                                                                           & $ 2.70\pm 0.21$   & $ 1.56\pm 0.10$  & 4176 \\
All (weighted)                             & $-$                                                                           & $ 3.45\pm 0.25$   & $ 1.47\pm 0.08$  & 3820 \\
\hline
\hline
\end{tabular}
\caption[high and low \hi mass $r_0$ and $\gamma$ values]{Measured values of $r_0$ and $\gamma$ for high and low \hi mass galaxies, high and low luminosity galaxies, high and low rotational velocity galaxies, and the full \hicat sample.}
\label{tab:r0gamma_mass}
\end{center}
\end{table*}

\clearpage

\clearpage

\clearpage

\clearpage

\clearpage

\clearpage

\clearpage

\clearpage

\clearpage

\clearpage

\clearpage

\clearpage

\clearpage

\clearpage

\label{lastpage}
\end{document}